\documentclass[structabstract]{aa}  
\usepackage[breaklinks]{hyperref}  
\usepackage[hyphenbreaks]{breakurl}
\usepackage{natbib,twoopt}

\usepackage{graphicx}
\usepackage{epstopdf}
\usepackage{natbib}
\usepackage{multirow}
\usepackage{xcolor}
\usepackage{rotating}
\usepackage{lscape}
\usepackage{longtable}
\usepackage{lipsum}

\makeatletter
  \newcommandtwoopt{\citeads}[3][][]{\href{http://adsabs.harvard.edu/abs/#3}%
    {\def\hyper@linkstart##1##2{}%
     \let\hyper@linkend\@empty\citealp[#1][#2]{#3}}}
  \newcommandtwoopt{\citepads}[3][][]{\href{http://adsabs.harvard.edu/abs/#3}%
    {\def\hyper@linkstart##1##2{}%
     \let\hyper@linkend\@empty\citep[#1][#2]{#3}}}
  \newcommandtwoopt{\citetads}[3][][]{\href{http://adsabs.harvard.edu/abs/#3}%
    {\def\hyper@linkstart##1##2{}%
     \let\hyper@linkend\@empty\citet[#1][#2]{#3}}}
  \newcommandtwoopt{\citeyearads}[3][][]%
    {\href{http://adsabs.harvard.edu/abs/#3}
    {\def\hyper@linkstart##1##2{}%
     \let\hyper@linkend\@empty\citeyear[#1][#2]{#3}}}
\makeatother

\usepackage{txfonts}
\bibpunct[]{(}{)}{;}{a}{}{,}
\begin{document}
   \title{The WFCAM multiwavelength Variable Star Catalog}

   \author{C. E. Ferreira Lopes$^{1}$ \and I.~D\'ek\'any$^{2,3}$ \and M.~Catelan$^{2,3}$ \and N. J. G.~Cross$^6$ \and R.~Angeloni$^{2,4,5}$  \and I. C. Le\~ao$^{1}$ \and J. R. De Medeiros$^{1}$}

   \institute{Departamento de F\'isica Te\'orica e Experimental, Universidade Federal do Rio Grande do Norte, 59072-970 Natal, RN, Brazil
         \and Instituto de Astrof\'isica, Pontificia Universidad Cat\'olica de Chile, Av. Vicu\~{n}a Mackenna 4860, 782-0436 Macul, Santiago, Chile
         \and Millennium Institute of Astrophysics, Santiago, Chile
         \and Centro de Astro-Ingenier\'{\i}a, Pontificia Universidad Cat\'olica de Chile, Santiago, Chile
         \and Max-Planck-Institut f\"{u}r Astronomie, K\"{o}nigstuhl 17, 69117 Heidelberg, Germany
         \and SUPA (Scottish Universities Physics Alliance) Wide-Field Astronomy Unit, Institute for Astronomy, School of Physics and Astronomy, University of Edinburgh, Royal
Observatory, Blackford Hill, Edinburgh EH9 3HJ, UK
             }

   \date{Received ; accepted}

 
  \abstract
   { Stellar variability in the near-infrared (NIR) remains largely unexplored. The exploitation of public science archives with data-mining methods offers a perspective for a time-domain exploration of the NIR sky.}
   {We perform a comprehensive search for stellar variability using the optical-NIR multiband photometric data in the public Calibration Database of the WFCAM Science Archive (WSA), with the aim of contributing to the general census of variable stars and of extending the current scarce inventory of accurate NIR light curves for a number of variable star classes.}
   {Standard data-mining methods were applied to extract and fine-tune time-series data from the WSA. We introduced new variability indices designed for multiband data with correlated sampling, and applied them for preselecting variable star candidates, i.e., light curves that are dominated by correlated variations, from noise-dominated ones. Preselection criteria were established by robust numerical tests for evaluating the response of variability indices to the colored noise characteristic of the data. We performed a period search using the string-length minimization method on an initial catalog of 6551 variable star candidates preselected by variability indices. Further frequency analysis was performed on positive candidates using three additional methods in combination, in order to cope with aliasing.}
   {We find 275 periodic variable stars and an additional 44 objects with suspected variability with uncertain periods or apparently aperiodic variation. Only 44 of these objects had been previously known, including 11 RR~Lyrae stars on the outskirts of the globular cluster M3 (NGC~5272). We provide a preliminary classification of the new variable stars that have well-measured light curves, but the variability types of a large number of objects remain ambiguous. We classify most of the new variables as contact binary stars, but we also find several pulsating stars, among which 34 are probably new field RR~Lyrae, and 3 are likely Cepheids. 
   We also identify 32 highly reddened variable objects close to previously known dark nebulae, suggesting that these are embedded
young stellar objects. We publish our results and all light curve data as the WFCAM Variable Star Catalog.
 }
   {}


   \maketitle


\section{Introduction}

Time-varying celestial phenomena in general represent one of the most substantial sources of astrophysical information, and their study has led to many fundamental discoveries in modern astronomy. Pulsating stars provide insight into to stellar interiors through asteroseismology \citep[e.g.,][]{gh12} and serve as standard candles \citep[e.g.,][]{aw12}, eclipsing binaries allow us to determine the most accurate stellar masses and radii \citep[e.g.,][]{jcea08}, and supernovae provide important means for estimating cosmological distances and probing the large-scale structure of our Universe \citep[e.g.,][]{area98,jtea03}~-- just to mention some classical scopes among the countless aspects of time-domain astronomy.

The ever-growing interest in various astronomical time-series data, as well as the tremendous development in astronomical instrumentation and automation during the past two decades, have been giving rise to several time-domain surveys of increasing scale. Wide-field shallow optical imaging surveys using small, dedicated telescope systems have been scanning the sky since the early 2000s with aims ranging from comprehensive stellar variability searches to exoplanet hunting, such as the Northern Sky Variability Survey (NSVS, \citealt{hoffman2009}), All Sky Automated Survey (ASAS, \citealt{2002AcA....52..397P}), Wide Angle Search for Planets (WASP, \citealt{2006PASP..118.1407P}), and the Hungarian Automated Telescope Network (HATnet, \citealt{2004PASP..116..266B}). The interest in transient events like microlensing have led to deeper, higher-resolution photometric campaigns such as the Massive Astrophysical Compact Halo Objects Survey (MACHO, \citealt{1993ASPC...43..291A}) and the Optical Gravitational Lensing Experiment (OGLE, \citealt{2003AcA....53..291U}), and to some ultra-wide surveys such as the Palomar Transient Factory (PTF, \citealt{2009AAS...21346901L}) or the Catalina Real-time Transient Survey \citep{adea09}. The development of imaging technology is leading toward deep all-sky surveys with seeing-limited resolution, and Pan-STARRS \citep{nkea02}, representing the first generation of such programs, is already operational. In the near future, even more ambitious programs, such as the Large Synoptic Survey Telescope (LSST, \citealt{2010SPIE.7733E..11K}) and Gaia \citep{2005ASPC..338....3P}, are planned to start monitoring the optical sky. These synoptic surveys are expected to be the new powerhouses of modern astronomy by providing a high data flow for a wide range of science applications, from the study of transients to understanding the dynamics of the Milky Way galaxy.

While optical synoptic surveys are getting wider and deeper, extending the systematic exploration of the variable sky toward other wavelengths, such as the infrared, is also indispensable, not only for a more complete understanding of the observed phenomena, but also for overcoming the problem of interstellar extinction. Infrared time-series photometry had long been constrained mostly to follow-up observations of known objects, but in recent years near-infrared (NIR) imagers have been converging to optical CCDs in terms of resolution and performance, hence wide-field surveys became feasible with some NIR instruments, such as the Wide-Field near-IR CAMera (WFCAM; \citealt{casali2007}) on the 3.8m United Kingdom Infrared Telescope (UKIRT), and the VISTA InfraRed CAMera (VIRCAM; \citealt{gdea06}) on the 4.1m Visible and Infrared Survey Telescope for Astronomy (VISTA), both hosting a variety of deep Galactic and extragalactic surveys. Nevertheless there are only a handful of wide-field NIR time-domain surveys that have more than a handful of observational epochs, with only the VISTA Variables in the V\'ia L\'actea survey \citep[VVV;][]{dmea10} being comparable to the optical ones in terms of both areal and time-domain coverage \citep[e.g.,][]{maea07,maea12}.

Besides the results from the large-scale surveys, valuable time-series data are also being generated by a variety of observational programs as sideproducts. Such data can well be quite heterogeneous and not always be complete completeness, since the original design of the data acquisition might have served various specific purposes. Nevertheless, multi-epoch data accumulated over a period of time from various observational programs using the same facility can hold vast unexploited information and represent a potential treasure trove for time-domain astronomy (see, e.g., the TAROT variable star catalog by \citealt{2007AJ....133.1470D}). An increasing number of observatories and projects realize the importance of standardizing their archives and incorporating them into the Virtual Observatory, allowing the community to exploit their data further once their proprietary periods have expired. There is already a wealth of fully public science-ready synoptic data from state-of-the-art instruments, which are accessible by standard data mining tools, waiting for analysis.

This paper is based on public time-domain data from the WFCAM Science Archive (WSA; \citealt{2008MNRAS.384..637H}). WFCAM was designed to be capable of carrying out ambitious large-scale survey programs such as the UKIRT Infrared Deep Sky Surveys \citep[UKIDSS;][]{lawrence2007}. The detector consists of four arrays of $2048\times2048$ pixels, arranged in a non-contiguous square pattern, providing a resolution of $0.4\arcsec$ per pixel. A contiguous image with an areal coverage of $0.78\,\rm{deg}^2$ can be constructed by a mosaic of four consecutive offset pointings. Other properties of the instrument are discussed in full detail by \cite{casali2007}. The standard UKIRT photometric system consists of the MKO $J$, $H$, and $K$ bands \citep[see][]{simons+02,tokunaga+02} complemented with the $Z$ and $Y$ filters. In addition, three narrow-band filters ($H_2$, $Br\gamma$, and $nbj$) are also available (see \citealt{hewett2006} for further details).

All WFCAM data produced by the UKIDSS surveys and other smaller campaigns and PI projects employing the instrument are processed by the VISTA Data Flow System (VDFS; \citealt{emerson2004}). VDFS was designed to initially handle UKIDSS, and includes both pipeline processing at the Cambridge Astronomical Survey Unit (CASU\footnote{\url{http://casu.ast.cam.ac.uk/surveys-projects/wfcam/technical/photometry}}) and further pipeline processing of the CASU products and digital curation at the WSA. Data for standard photometric zero-point calibration of and color term determination for the WFCAM filters are taken regularly and are also fully processed and archived by the VDFS in the same way as the survey data. During the several years of operation of WFCAM, a large amount of fully processed, high-quality, multiband photometric data have been collected during the observations of calibration fields, which are publicly available at the WSA. Since the majority of these fields have been visited several times over many years, these data provide an excellent opportunity for studying of variable stars in the NIR.

In this paper, we perform a comprehensive stellar variability analysis of the public WFCAM Calibration (WFCAMCAL) Database (release 08B) and present the photometric data and characteristics of the identified variable stars as the WFCAM Variable Star Catalog, version~1 (WVSC1). The paper is structured as follows. In Sect.~\ref{wfcamcal}, we present the database and its characteristics from a data miner's point of view and describe the primary source selection procedure. In Sect.~\ref{varindices}, we introduce and discuss a set of variability indices designed for synoptic data with correlated sampling, and we employ these indices as a first selection of candidate variable sources in the WFCAM data. We present a frequency analysis of the candidate variable sources in Sect.~4. In Sect. 5, we present the WVSC1 by describing its structure and discussing its variable star content. Finally, in Sect. 6, we draw our conclusions and discuss some future perspectives.


\section{The WFCAMCAL database}
\label{wfcamcal}

WFCAM on-sky image data are pipeline-processed by the VDFS, which incorporates all data reduction steps from image processing to photometry. The various processing steps are described fully by \cite{emerson2004}, and accordingly we limit ourselves to discussing some key aspects. Individual exposures go through the standard preprocessing steps, such as flat fielding and dark subtraction. As is common practice in the NIR, science frames are composed of a set of dithered images, i.e. subsequent exposures taken in step patterns of several arcseconds, in order to allow for the removal of bad detector pixels and to image the structure of the atmospheric IR foreground for its subtraction before stacking the dither sequence. These so-called \textit{detector frame} stacks are the final image products of calibration field observations, but we note that some WFCAM survey images are combined further to form contiguous mosaic images also known as \textit{tiles}. Deep stack images are also produced, in order to push the limiting magnitudes and measure the fluxes of faint sources.

Further steps in the reduction include source extraction and aperture photometry using a set of small circular apertures with radii of $0.5$, $\sqrt{2}/2$, $1$, $\sqrt{2}$, and $2$ arcseconds, in order to maximize the signal-to-noise ratio and remedy systematics due to source crowding \citep{2009MNRAS.399.1730C}. Flux loss in the wings of the point spread functions (PSF's) is corrected for \citep{2004SPIE.5493..411I}. Each identified source of the catalogs is classified based on the shape of its PSF, and various quality flags are also assigned. Astrometric calibration is performed using 2MASS \citep{msea06} stars as reference, and positions have a typical accuracy of $0.1 \arcsec$. Magnitudes are corrected for distortion effects and are zero-point-calibrated on a frame-by-frame basis using a many of secondary standard stars from the 2MASS catalog \citep{msea06}. This involves colour corrections from the 2MASS $JHKs$ to WFCAM $ZYJHK$. The calibration process is described in detail in \citep{Hodgkin2009}. Many calibration fields (which form the basis of this paper) are also observed repeatedly under photometric conditions, in order to provide standard star measurements for the calibrations of the $Z$, $Y$, and narrow-band magnitudes, as well as to provide data for the color-term determination for transforming of 2MASS magnitudes into the WFCAM photometric system. The accuracy of the magnitude zero-points of these fields is a few percent. We note that all magnitude data in the WSA and in our study alike are on the WFCAM magnitude system.

The calibrated source catalogs undergo further curation steps at the WSA, including quality control, source merging, and both internal and external cross-matching. Enhanced image products (e.g., deep stacks) are also created. Final science-ready data are ingested into a relational database, which offers various server-side data management tools. Data can be queried by using the Structured Query Language (SQL). The design of the WSA, the details of the data curation procedures and the layout of the database are described in great detail in \citet{2008MNRAS.384..637H} and \citet{2009MNRAS.399.1730C}. In the following, we highlight some other properties of the data in the WFCAMCAL archive, with an emphasis on some important details for the time-series analysis.

\subsection{Data characteristics}
\label{datachar}

\begin{figure}[t]
 \includegraphics[width=0.5\textwidth]{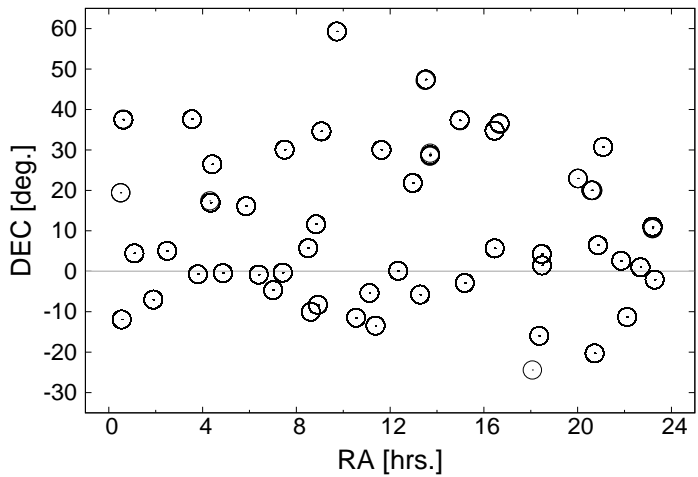}
 \caption{Celestial distribution of our initial database of stellar sources from the WFCAMCAL08B archive.}
 \label{fig_overview01}
 \end{figure}

The WFCAMCAL archive's data release 08B contains data from 52 individual pointings from both the northern and southern hemispheres, spread over nearly half of the sky. These calibration fields are located between declinations of $+59\fdg62$ and $-24\fdg73$, distributed over the full range in right ascension in order to provide standard star data year-round. The positions of the pointings are shown in Figure~\ref{fig_overview01}. Each pointing consists of a non-contiguous area covered by the four WFCAM chips, covering $\sim0.05$ square degrees each. The total area covered by all standard star fields is $\sim10.4$ square degrees.

\begin{table}[htbp]
\caption{Morphological classification of sources in the WFCAMCAL archive}
\centering                     
\begin{tabular}{clr@{,}l}     
\hline\hline
Class & Description & \multicolumn{2}{r}{Number of sources}  \\ 
\hline               
  $-1$ & Star    &    318 & 995 \\  
  $-2$ & Probable star    &    33 & 188 \\  
  $+1$ & Galaxy    &    205 & 352 \\  
  $-3$ & Probable galaxy    &    2 & 345 \\  
  $0$ & Noise    &    10 & 377 \\  
\hline
\end{tabular}
\label{tab_class}
\end{table}

\begin{figure}[t]
 \includegraphics[width=0.5\textwidth]{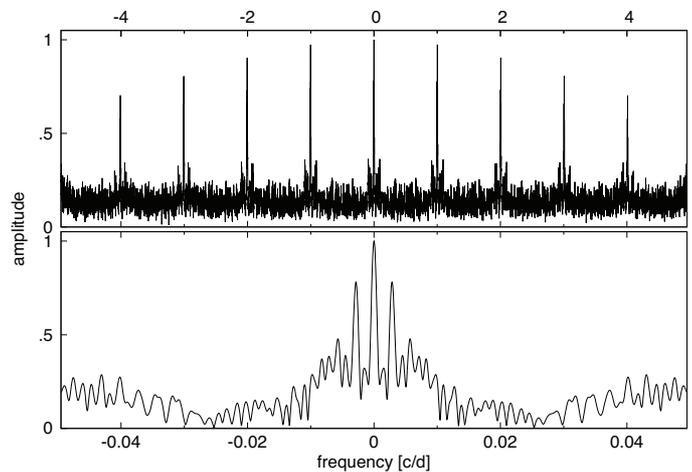}
 \caption{Spectral window of a typical well-sampled $K$-band light curve from the WFCAMCAL database, showing two different frequency ranges.}
 \label{spwin}
 \end{figure}

The majority of the fields are observed repeatedly, with a rather irregular sampling. A field is usually visited at most a few times during a night, with the visits separated by up to a few hours. During each visit, the field is usually (but not necessarily) observed through the $JHK$ or the $ZYJHK$ filter set, and occasionally through the narrow-band filters as well, all within a few minutes. A certain field is usually observed again within a few days, although longer time gaps are common, and of course large seasonal gaps are also present in the data set. The total baseline of the time series is largely field-dependent and varies from a few months up to three years. The time sampling (cadence) in a single passband can be considered to be quasi-stochastic with rather irregular gaps, which is a favorable scenario for detecting of periodic signals over a wide range of periods. Figure~\ref{spwin} shows the spectral window of a typical one among the well-sampled $K$-band light curves in the database, showing that only one-day and one-year aliases are significant. However, the sampling in different passbands is very strongly correlated.

The VDFS performs an accurate and deep source extraction procedure on deep stack images, which, in the case of the WFCAMCAL database, is merged from the seven best-seeing frames to reduce any problems with blending \citep{2009MNRAS.399.1730C}. This curation step gives rise to database entities known as \textit{Sources}. Positional cross-matches between these \textit{Sources} and detections from different single detector frame stacks are then performed for each frame set (i.e., images with the same pointing). The prescriptions for this cross-matching procedure and its limitations are given by \citet[][, see their Sect.~3.4.2]{2008MNRAS.384..637H}. Each \textit{Source} of the database contains only partial information from a set of source detections positionally matched with each other, but is related to a set of pointers that track all the attributes of the corresponding detections. We note that after cross-matching, the VDFS also recalibrates the individual epochs after comparison to the deep stack to improve the light curves. The resulting source-merged catalog is stored in WSA's relational database, which in the case of WFCAMCAL data, follows a synoptic database model designed for observations with correlated sampling \citep{2009MNRAS.399.1730C}. Since the observations of the standard star fields through different filters are always taken within a time interval that is several orders of magnitude shorter than the interval between two successive observation batches (i.e., visits of the same field, see above), they can be considered to be virtually at the same epoch. Individual source detections in different filters for each epoch (i.e., filter sequence) are therefore merged and stored in a database entity called synoptic source, and these are linked to  \textit{Sources} (see above). Various metadata of the time series formed by the detections associated to each \textit{Source} are also provided by WSA as the attributes of the variability table, providing the best aperture for a given source, together with basic statistics on the magnitude distribution, time sampling, etc. In passing we note that this table also provides a basic assessment on the probability that a source is variable in each band, based on the comparison of the rms scatter of the time series to its expected value from a simple noise model \citep[see][]{2009MNRAS.399.1730C}. Although these pieces of information can provide useful guidance for the user, we opted not to use these attributes as criteria in our source selection, since our aim is to perform a more profound variability search in the data.

\subsection{Initial source selection}
\label{sourcesel}

Data were retrieved in a two-step procedure via WSA's freeform SQL query facility.\footnote{\url{http://surveys.roe.ac.uk:8080/wsa/SQL\_form.jsp}} We queried all \textit{Sources} that were classified as a star or probable star based on the PSF statistics of the merged source, having at least ten unflagged epochs in either of the eight filter passbands. 

\begin{table}[]
\caption{Number of sources, average total baseline ($\langle{T}_{\rm tot}\rangle$, in days), and average number of epochs $\langle{N}_{\rm ep}\rangle $ in each filter, in our initial database.}
\centering                          
\begin{tabular}{l r@{,}l c c}        
\hline\hline                 
\multicolumn{1}{c}{Filter}  &
\multicolumn{2}{c}{${N}$} & 
\multicolumn{1}{c}{$\langle{T}_{\rm tot}\rangle$} & 
\multicolumn{1}{c}{$\langle{N}_{\rm ep}\rangle $} \\    
\hline                        
   $Z$     & 204 & 245 & 1033 & 51 \\
   $Y$     & 212 & 334 & 1061 & 56 \\      
   $J$      & 212 & 717 & 1058 & 58 \\
   $H$     & 212 & 169 & 1074 & 61 \\
   $K$     & 201 & 645 & 1065 & 60 \\ 
\hline
   $H_2$  & 110 & 587  & 429  & 4  \\ 
   $Br\gamma$  & 130 & 307  & 258  & 2  \\ 
   $nbj$ & 68 & 048   & 120  & 3  \\ 
\hline                                   
\end{tabular}
\label{tspanep}
\end{table}

\begin{figure}
\includegraphics[width=0.5\textwidth]{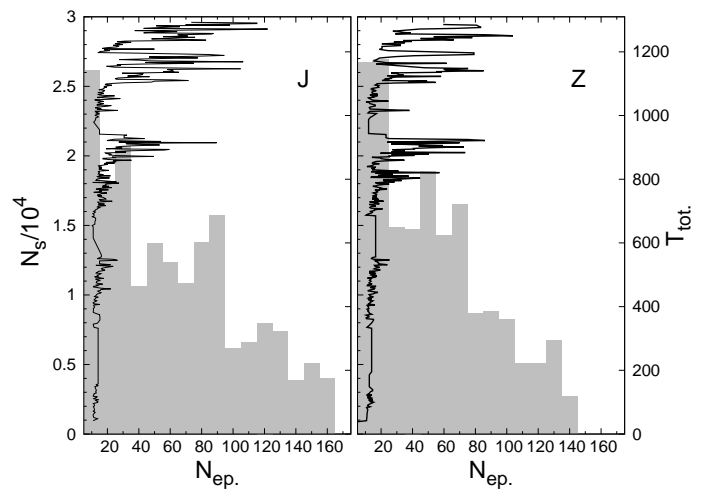}
\caption{Histograms showing the number of sources ($N_{\rm s}$) as a function of the number of epochs ($N_{\rm ep.}$) for the queried $J$- (left) and $Z$-band (right) light curves with a bin size of 10 days. The distribution corresponding to the other broad bands are very similar. The solid curves show the average total baseline length as a function of the number of epochs for the queried light curves in the same filters with a bin width of 2 days.}
\label{fig_overview02}
 \end{figure}

Our selection resulted in an \textit{initial database} of $216, \!722$ sources containing their identifiers and general attributes such as positions, basic flux, and sampling statistics. Figure~\ref{fig_overview02} shows histograms of the queried light curves and the average total baseline length as a function of the number of epochs, while Tables~\ref{tab_class} and \ref{tspanep} summarize some basic statistics of the queried data. The sampling of the time domain is rather heterogeneous, thus we must emphasize that the WFCAMCAL archive should not be considered as a synoptic survey database, for its completeness is highly varying from field to field; accordingly, care must be taken with any statistical interpretation based on these data. However, there are a large number of light curves with very good sampling that have $\sim 100$ data points in three to band broad-band filters, with a total baseline of up to 3 years. Since the number of data points in the narrow-band filters is generally very low, we do not use these measurements for the variability search, but they are provided (if available) for the detected variable sources in our catalog.

With the result of our first selection query at hand, we queried the complete light curves of each source by linking their entries in the \textit{Source} table to the attributes of the \textit{Synoptic\,Source} table. This procedure requires the generation of a temporary SQL table provided by the user containing the identifiers of the \textit{Sources} from the first query. We note that some data, such as Julian Dates and zero-point errors, are stored in different database objects like the \textit{Multiframe} table, which require the execution of further table joins. We selected magnitudes with sufficiently small error bits ($<128$), rejecting highly saturated measurements and data affected by various severe defects (e.g., bad pixels, poor flat-field region, detection close to
frame boundary, etc.). 
 
\section{Broad selection by variability indices}
\label{varindices}

A key characteristic of the time-series data in the WFCAMCAL database is that, owing to the observation strategy, their sampling is strongly correlated in the different filters (see Sect.~\ref{datachar}). The characteristic time lag between data points in different filters in an observing sequence (i.e., a standard visit of a calibration field) is not longer than a few minutes, which is orders of magnitude shorter than the typical time gap between such batches of data and also than the time scales of stellar variability at amplitudes that are typically recoverable by the present sampling and accuracy. This property of the data enables us to search for stellar variability through correlations between the temporal flux changes at different wavelengths.

In our approach to stellar variability searching, we make the following general assumptions: (i)~intrinsic stellar variability is typically identifiable in a wavelength range that is wider than our broad-band filters (i.e., in more than one filter); (ii)~it is sufficiently phase-locked at two close wavelengths, thus flux variations in neighboring wavebands will be correlated; and (iii)~non-intrinsic variations will be typically stochastic. In point (iii) we also implicitly assume that wavelength-correlated systematics of instrumental and atmospheric origin, or due to possible data reduction anomalies, have amplitudes that are small to be enough comparable to those provided by stochastic variations. The most important deviation from this assumption for our data is due to the temporal saturation of bright objects, which can, however, be easily distinguished from stellar variability.

A commonly used approach to identifing stellar variability through correlations in flux changes is by the Welch-Stetson index ($I_{\rm WS}$; \citealt{1993AJ....105.1813W}; \citealt{1996PASP..108..851S}). The idea behind this index is to separate stochastic variations from systematic trends by measuring the correlations in the deviations from the mean of data points that are located sufficiently close in time. It is defined as

 \begin{equation}
    I_{\rm WS} = \sqrt[ ]{\frac{1}{n(n - 1)}}\sum_{i=1}^{n}\left(\delta b_i \delta v_i \right)\,,
 \label{eqstet}    
 \end{equation}

\noindent where

 \begin{equation}
   \label{dflux}
    \delta b_i = \frac{b_i - \bar{b}}{\sigma_{b,i}}\,,~\delta v_i = \frac{v_i - \bar{v}}{\sigma_{v,i}}\,,
 \end{equation}

\noindent with $b_i$, $v_i$, and $\sigma_{b,i}, \sigma_{v,i}$ denoting magnitudes and their errors (the latter computed following the prescriptions by \citealt{pbs81}), respectively, taken on a time scale that is much shorter than that of the intrinsic stellar variations of interest. These magnitudes and errors can represent either successive monochromatic measurements or data points taken in two different filters. We also note that $\bar{b},\bar{v}$ are weighted averages. Finally, $n$ denotes the number of epochs, by which we also refer to the number of short time slots that contain subsequently taken measurements in the case of non-simultaneously observed data with strongly correlated sampling. The $I_{\rm WS}$ index is found to be significantly more sensitive than the ``traditional'' $\chi^2$-test for single variance, which uses the magnitude-rms scatter distribution of the data as a predictor (see, e.g., \citealt{2002AcA....52..397P}).

\subsection{Extension of $I_{WS}$ to multiband datasets}\label{genws}

According to the definition given in Eq.~(\ref{eqstet}), $I_{\rm WS}$ is limited to pairwise comparisons of fluxes. When it is applied its application to panchromatic data with measurements taken in several wavebands with correlated sampling, one would need to define some conversion of the pairwise indices into a single quantity. To accomplish this, we introduce the following modification to $I_{\rm WS}$, for quantifying panchromatic flux correlations (pfc):

\begin{equation}
     I_{\rm pfc}^{(2)} = \sqrt{\frac{(n_{2}-2)!}{n_{2}!}}\sum_{i=1}^{n}\left\{ \sum_{j=1}^{m-1}\left[ \sum_{k=j+1}^{m}\left( \delta u_{ij} \delta u_{ik} \right) \right] \right\}\,,
\label{eq_pfc2}
\end{equation}

\noindent where $n$ is the number of epochs, $m$ is the number of wavebands, $u_{ij}$ are the flux measurements, $\delta u_{ij}$ is defined by Eq.~(\ref{dflux}) for each filter, and $n_{2} = n\cdot m!/[2!(m-2)!]$. We note that, for $m=2$, $I_{\rm pfc}^{(2)} = I_{\rm WS}$.

While Eq.~(\ref{dflux}) still measures variability through pairwise correlations of relative fluxes, we can generalize it for quasi-simultaneously measured batches of three data points (in case of $m \ge 3$) by

   \begin{equation}
     I_{\rm pfc}^{(3)} = \sqrt{\frac{(n_{3}-3)!}{n_{3}!}}\sum_{i=1}^{n}\left\{  \sum_{j=1}^{m-2}\left[ \sum_{k=j+1}^{m-1}\left( \sum_{l=k+1}^{m}\Lambda_{ijkl}^{(3)} \left|  \delta u_{ij}  \delta u_{ik}  \delta u_{il}  \right| \right) \right] \right\}\,,
   \label{eq_pfc3}     
   \end{equation}
   
\noindent where $n_{3} = n\cdot m!/[3!(m-3)!]$, and the $\Lambda$ factor is defined as

\begin{equation}
      \Lambda_{ijkl}^{(3)} = \left\{ \begin{array}{ll}
      +1 & \qquad \mbox{if \,\, $\delta u_{ij} > 0, \,\,  \delta u_{ik} > 0, \,\, \delta u_{il} > 0$ };\\
      +1 & \qquad \mbox{if \,\, $\delta u_{ij} < 0, \,\,  \delta u_{ik} < 0, \,\,  \delta u_{il} < 0$ };\\
      -1 & \qquad \mbox{otherwise}.\end{array} \right. 
\label{eq_lambda3}    
\end{equation}

\noindent We note that we introduced $\Lambda^{(3)}$ to give the proper sign to the product in Eq.~(\ref{eq_pfc3}). 

Finally, in the case of quasi-simultaneously measured batches of $s$ data points (for $m \ge s$), our variability index takes the following general form:
  
\begin{equation}
      I_{\rm pfc}^{(s)} = \! \sqrt{\frac{(n_{s}-s)!}{n_{s}!}}\sum_{i=1}^{n}\left [  \sum_{j_1=1}^{m-(s-1)}\!\!\!\cdots \left ( \sum_{j_s=j_{(s-1)}+1}^{m} \!\!\! \Lambda_{ij_1\cdots j_{s}}^{(s)}  \left| \delta u_{ij_1} \cdots \delta u_{ij_s}  \right| \right ) \right ],
   \label{eq_pfcgen}     
\end{equation}   

\noindent where $n_s = n\cdot m!/[s!(m-s)!]$, and the $\Lambda^{(s)}$ correction factor is

\begin{equation}
      \Lambda_{ij_1\cdots j_{s}}^{(s)} = \left\{ \begin{array}{ll}
      +1 & \qquad \mbox{if \,\, $\delta u_{ij_1} > 0, \,\, \cdots \,, \, \delta u_{im} > 0$ };\\
      +1 & \qquad \mbox{if \,\, $\delta u_{ij_1} < 0, \,\, \cdots \,, \, \delta u_{im} < 0$ };\\
      -1 & \qquad \mbox{otherwise}.\end{array} \right. 
   \label{eq_feps}    
\end{equation} 


\noindent Clearly, these indices set increasingly strict constraints on the presence of variability with increasing order $s$.

\subsection{An alternative variability index}\label{newvar}

While the $I_{\rm pfc}^{(s)}$ index is quite robust, in the sense that it is weighted with the individual errors, it can be insensitive to true variable stars for one or several substantially outlying data points when incorrect error estimates are present. Indeed, this index may even introduce false variability candidates if these outliers are correlated in two or more bands. Although such situations might seem rare, NIR data in particular can present us with these cases quite frequently, particularly in the case of bright stars. Since the sky foreground emitted by the atmosphere is highly variable in the NIR, it causes a highly time-varying saturation limit, which can affect large parts of otherwise highly accurate time-series data for bright stars with substantial outliers having very small formal error estimates. In case of correlated sampling, these outliers will probably be correlated between different filters, leading to a spurious impact upon the $I_{\rm pfc}^{(s)}$ index.

To alleviate the effect of such anomalous outliers, we introduce an alternative variability index that is similar to $I_{\rm pfc}^{(s)}$, but does not depend on the actual value of the flux deviations from the mean. This is simply obtained by keeping only the $\Lambda$ function (Eq.~\ref{eq_feps}) in the sum that appears in Eq.~(\ref{eq_pfcgen}). Thus, the flux-independent (fi) version of $I_{\rm pfc}^{(2)}$ is defined as
   
   \begin{equation}
     2 \cdot I_{\rm fi}^{(2)} -1 = \frac{1}{n_{2}}\sum_{i=1}^{n}\left [ \sum_{j=1}^{m-1}\left (\sum_{k=j+1}^m \Lambda_{ijk}^{(2)} \right ) \right ]\ ,
	 \label{eq_fi2}     
   \end{equation}

\noindent with $\Lambda_{ijk}^{(2)}$ defined as   

\begin{equation}
      \Lambda_{ijk}^{(2)} = \left\{ \begin{array}{ll}
      +1 & \qquad \mbox{if \,\, $\delta u_{ij} > 0, \,\,  \delta u_{ik} > 0$ };\\
      +1 & \qquad \mbox{if \,\, $\delta u_{ij} < 0, \,\,  \delta u_{ik} < 0$ };\\
      -1 & \qquad \mbox{otherwise}.\end{array} \right.  
\label{eq_lambda2}    
\end{equation} 

\noindent In this expression, the $\delta u$'s have the same meaning as before (e.g., Eq.~\ref{eq_pfc3}).  
   
The righthand side of Eq.~(\ref{eq_fi2}) thus gives the difference between the number of positive and negative terms in $I_{\rm pfc}^{(s)}$, so it can only take a number of discrete values (depending on the values of $n$ and $m$). We note that, with the coefficients included on the lefthand side of Eq.~(\ref{eq_fi2}), $I_{\rm fi}^{(2)}$ will always have absolute values between 0 and 1, analogously to a probability measure.

For higher orders $s$, $I_{\rm fi}^{(s)}$ is defined similarly to Eq.~(\ref{eq_pfcgen}):

   \begin{equation}
      2 \cdot I_{\rm fi}^{(s)} -1 = \frac{1}{n_{s}}\sum_{i=1}^{n}\left [  \sum_{j_1=1}^{m-(s-1)}\cdots \left ( \sum_{j_s=j_{(s-1)+1}}^m \Lambda_{ij_1\cdots j_{s}}^{(s)} \right ) \right ]\,,
   \label{eq_figen}     
   \end{equation}   

\noindent with $0 \leq I_{\rm fi}^{(s)} \leq +1$, and where $\Lambda_{ij_1\cdots j_s}^{(s)}$ is defined as in Eq.~(\ref{eq_feps}).  

Thus, the index $I_{\rm fi}^{(s)}$ represents a problem of combining signals whose function $\Lambda_{ij_1\cdots j_{s}}^{(s)}$ (Eq.~\ref{eq_feps}) assumes the values $+1$ or $-1$, depending on the sign of the correlation. The total number of possible combinations of signals in a set of $s$ measurements, so that each group of signals is different from the other, is given by 

   \begin{equation}
     A_{s} = s^{2}. 
   \label{prob001}     
   \end{equation}

\noindent For the index $I_{\rm fi}^{(2)}$, for instance, we have four possible configurations, namely $(+ +, + -, - +, - -)$. According to elementary probability theory, in the case of statistically independent events, the probability that a given event will occur is obtained by dividing the number of events of the given type by the total number of possible events. For the $I_{\rm fi}^{(s)}$ index of any order, the desired events will be those in which all signals are either positive or negative, so that, irrespective of the value of $s$, the number of events desired always equals two, and the number of possible events is given by Eq.(~\ref{prob001}). In this way, the general expression that determines the probability value of a random event leading to a positive $I_{\rm fi}^{(s)}$ index is given by

   \begin{equation}
     P_{s} = \frac{2}{s^{2}}, 
   \label{prob002}     
   \end{equation}

\noindent given that there are $s^2$ events in total, but only two produce a positive $I_{\rm fi}^{(s)}$ value. 
   
\begin{figure}[t]
\centering
\includegraphics[width=9cm]{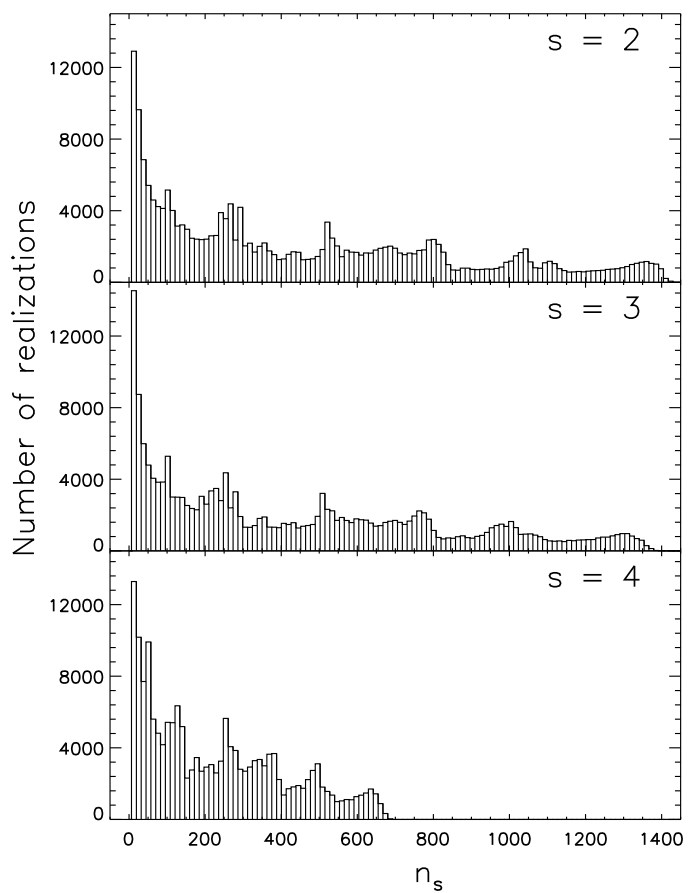}
\caption{Histograms showing the number of combinations $n_{s}$ in the simulations described in the text. Since the $I_{\rm fi}^{(s)}$ index is computed in a manner similar to that used to compute the $I_{\rm pfc}^{(s)}$ index, the number of combinations used is the same for both. }
\label{fig_hist_pfc}
\end{figure}

\subsection{Numerical tests}
\label{simulation}

To establish robust criteria for selecting of variable star candidates, we evaluated the responses of the $I_{\rm pfc}^{(s)}$ and $I_{\rm fi}^{(s)}$ indices to statistical fluctuations. We generated a large number of test time-series sequences by shuffling the  times (``bootstrapping'') of the WFCAMCAL data proper. By following this approach, we are able to keep part of the correlated nature of the noise intrinsic to the data, as opposed to numerical tests based on pure Gaussian noise. Figure~\ref{fig_hist_pfc} shows the histogram of the number of test time-series sequences as a function of the $n_s$ number of terms appearing in the variability indices (see Eqs.~\ref{eq_pfcgen}, \ref{eq_figen}) for various $s$ values. In total, $200,\!000$ realizations were performed to our tests.

\begin{figure*}[t]
\centering
  \includegraphics[width=\textwidth]{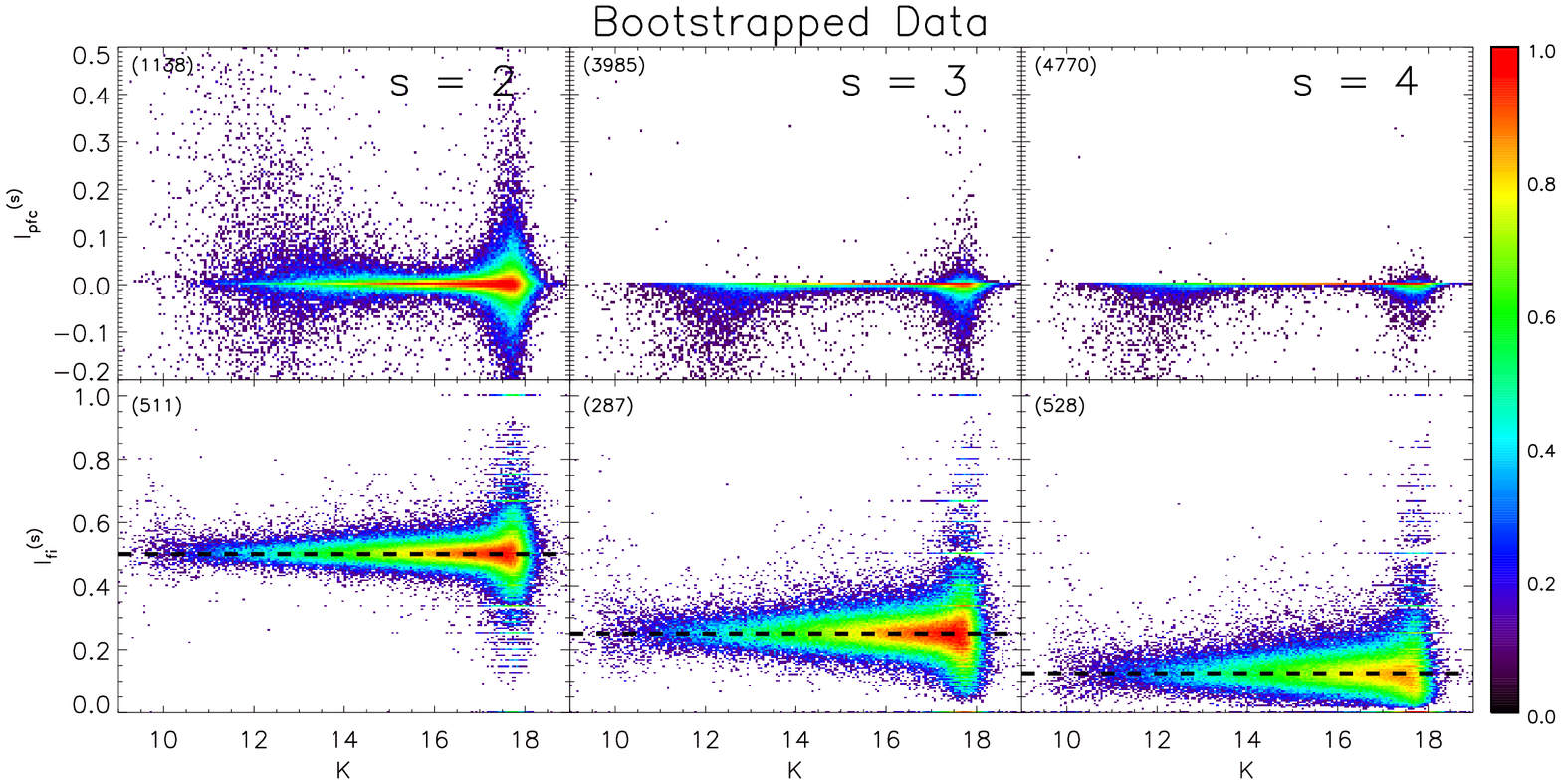}
  
  \includegraphics[width=\textwidth]{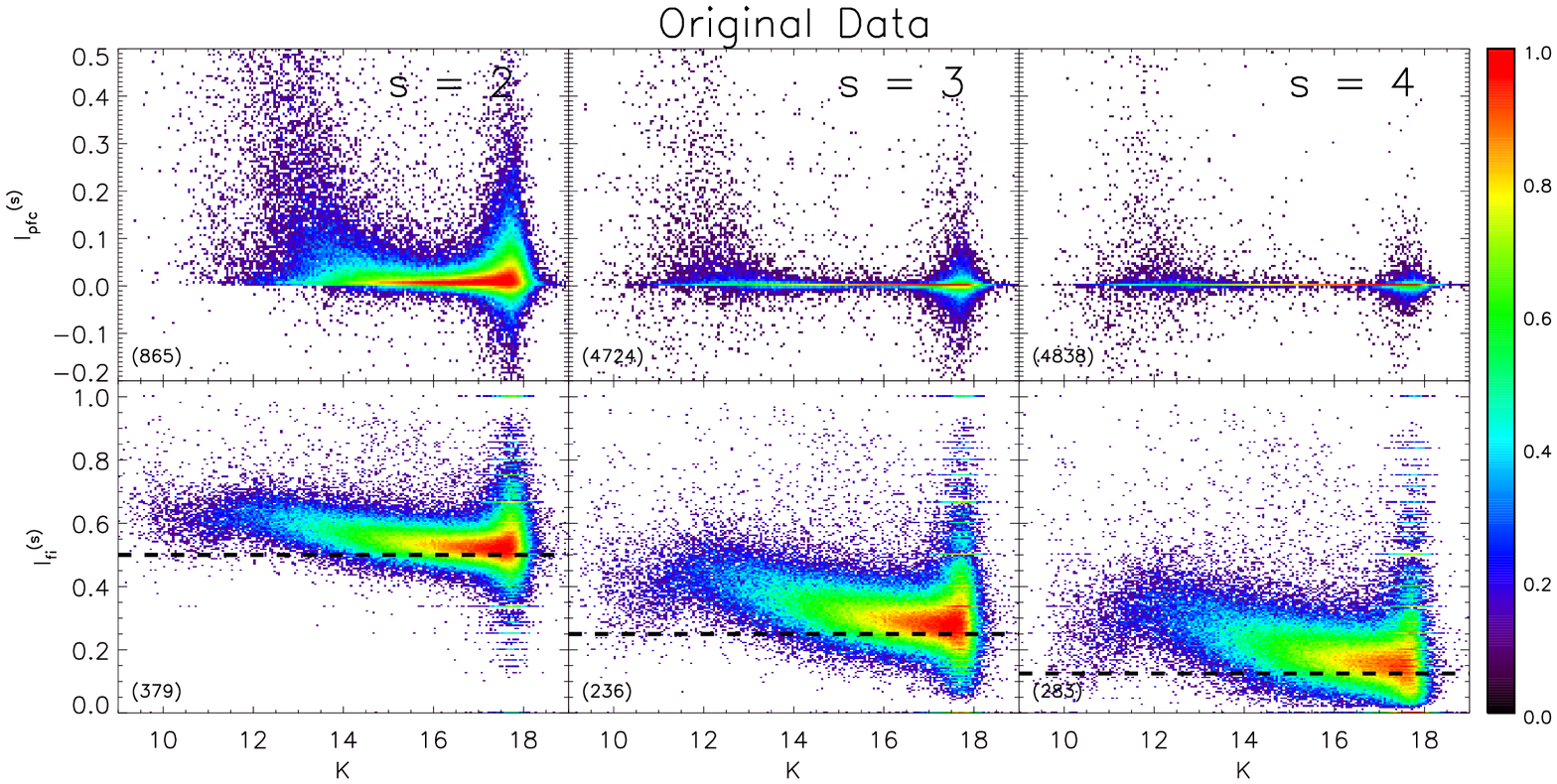}
    
  \caption{Variability indices versus the $K$-band magnitude, for the simulations (top two rows) and the actual data (bottom two rows), for three values of $s$, namely 2 (left), 3 (middle), and 4 (right). }
 \label{fig_indices}
\end{figure*}

Figure~\ref{fig_indices} shows the distribution of the variability indices as a function of the average apparent brightness of the sources for both the bootstrapped and the corresponding original data. The excess of high values for bright sources, which is particularly evident for $s = 2$, is due to temporal saturation, which usually happens in more than one waveband at the same time. In this bright regime, as expected (Sect.~\ref{newvar}), saturation affects the $I_{\rm pfc}^{(s)}$ index much more frequently than it does $I_{\rm fi}^{(s)}$. At the faint end, however, a number of effects can also be recognized. First, the quantized nature of the $I_{\rm fi}^{(s)}$ index (as opposed to the $I_{\rm pfc}^{(s)}$ index; see Sect.~\ref{varindices}) becomes pronounced in the distribution, owing to the higher relative frequency of sources with but a few epochs (typically $n_{s} \lesssim 20$) among the faint stars. (These are often not detected if the atmospheric foreground flux is too high.) Second, there is also an excess of sources with high index values at the faint end, which is due to the much lesser amount of data for these fainter sources, which makes the variability indices more sensitive to statistical fluctuations and systematics.
The general trend in the distributions of the $I_{\rm pfc}^{(s)}$ and $I_{\rm fi}^{(s)}$ are also rather different. Because it is sensitive only to the consistency in the direction of the flux changes, the $I_{\rm fi}^{(s)}$ is equally responsive to contaminating systematics of different amplitudes. This is reflected by both the slight increase in the main locus of the distribution of this index as a function of magnitude toward the bright end (see Fig.~\ref{fig_indices}, lower panels) and the enhanced (false) response in the faint end, as discussed earlier. These are both caused by the increasing dominance of correlated noise over photon noise. We note that this is not shown by the bootstrapped data at the bright end, since the points of distinct {\em pawprint} batches were shuffled, and thus the correlations between them were mostly lost, while this is alleviated by the small number of points per light curve at the faint end.

Also noteworthy is that the simulated $I_{\rm fi}^{(s)}$ distributions that are shown in Figure~\ref{fig_indices} are centered on the theoretically expected values (see Sect.~\ref{varindices}), as given by Eq.~(\ref{prob002}), namely $0.5$ in the case of $s=2$, $0.25$ in the case of $s=3$, and $0.125$ for $s=4$. Moreover, the simulation results for the $I_{\rm pfc}^{(s)}$ index show that the higher the order of the index, the lower the dispersion around $I_{\rm pfc}^{(s)} = 0$; however, a pronounced level of scatter remains at both the bright and faint ends of the distribution. In addition, there is a symmetry between scatterings for the index $I_{\rm pfc}^{(2)}$ and asymmetric scattering for $I_{\rm pfc}^{(3)}$ and $I_{\rm pfc}^{(4)}$, both of which primarily scatter toward negative values, particularly at the bright end. This happens because the number of possible configurations of signs that lead to a $\Lambda = +1$ (Eq.~\ref{eq_feps}) is always equal to two, whereas the number of possible configurations that lead to a $\Lambda = -1$ is instead given by $ s^{2}-2 $. In this way, for $s = 2$, we have a symmetrical distribution of configurations of signs, whereas for $ s> 2$ the asymmetry in relation to negative values grows as $s^{2}-2$, as seen in the actual simulations.

\begin{figure*}[t]
  \includegraphics[width=\textwidth]{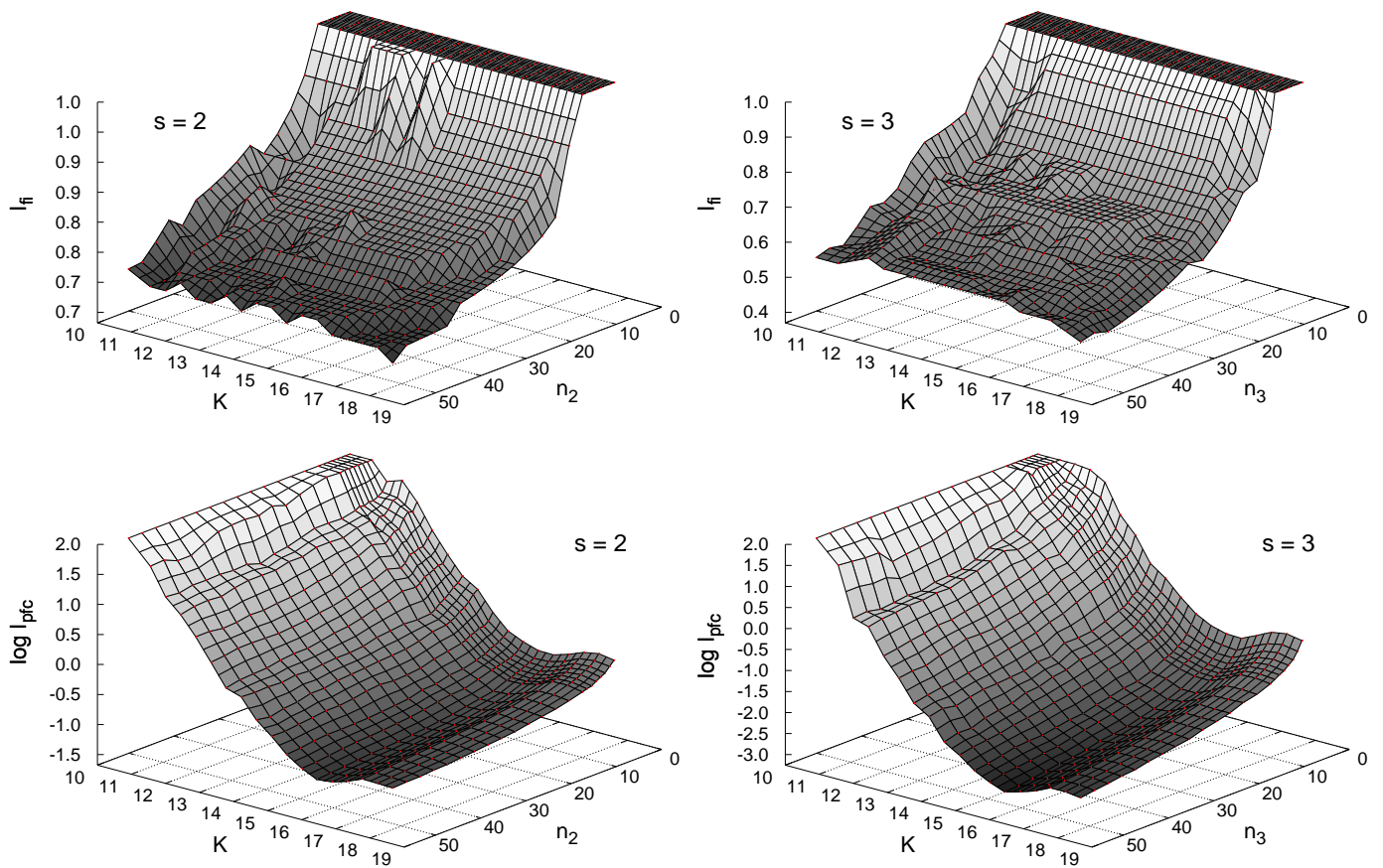}
  \caption{As in Figure~\ref{fig_indices}, but for the cutoff surfaces used to select our target list, and also adding $n_s$ as an independent variable. These surfaces set apart, at the $0.5\%$ significance level, instances that are compatible with random noise (low values of the indices) from those that are compatible with coherent signal being present in the different passbands (high values of the indices).}
 \label{cutoff_surface}
\end{figure*}

We estimated the significance levels of our variability indices based on their cumulative number density distributions obtained from our bootstrapped data set. In other words, we used our bootstrap simulations to set the signals apart that are compatible with pure random noise from the signals that indicate correlated variability in the different bandpasses. We expect these estimates to be highly dependent on the number of epochs $n$ and the brightness (see above); accordingly, we obtained separate estimates for data within narrow ranges of $n$ and $K$-band average magnitude, with values in between being obtained by interpolation. Figure~\ref{cutoff_surface} shows the result. In this figure, we show surfaces, in $({\rm index}, n_s, K)$ space that set random noise apart from correlated signals at the $0.5\%$ significance level, according to our two variability indices, namely $I_{\rm pfc}^{(s)}$ (top row) and $I_{\rm fi}^{(s)}$ (bottom row). As can be seen from Figure~\ref{cutoff_surface}, the corresponding cutoff values of the $I_{\rm fi}^{(s)}$ variability index show a relatively weak dependence on the magnitude and a strong dependence on $n_{s}$, whereas the $I_{\rm pfc}^{(s)}$ index depends strongly on both of these quantities. This is mainly attributed to the fact that the mean values of the formal errors (estimated from photon noise only), to which $I_{\rm pfc}^{(s)}$ is very sensitive, increasingly underestimate the global scatter of the light curves toward lower magnitudes in this dataset.

\subsection{Initial selection of variable star candidates}
\label{sec_selecvar}

We next computed the variability indices introduced in Sect.~\ref{varindices} for all sources in our \emph{initial database}. As many as $99.3\%$ of these sources had quasi-simultaneous observations in at least two filters, while $91.3\%$ and $81.3\%$ of them were observed in at least three and four filters, respectively. For each variability index, we selected all sources above its empirical, sampling, and magnitude-dependent 0.5\% significance level described in Sect. 3.3. We considered a source as a candidate variable star if either value of its variability indices was significant. This procedure resulted in our {\em initial catalog} of 6651 candidate variable stars.

\section{Frequency analysis}  \label{secperiod}

There are many methods available for computing periods in unevenly spaced time-series data, based mainly on Fourier analysis, information theory, and statistical techniques, among others \citep[see, e.g.,][, for a review]{mt04}. Some of these methods are based on the fact that the phase diagram of the light curve (also known as simply ``phased light curve'') is smoothest when it is visualized using its real period. They transform the set of data by folding within the phase interval $0 \le \varphi(i) < 1$, which is defined by the following expression: 

  \begin{equation}
     \varphi(i) = \frac{t(i) - t_{0}}{P}  - {\rm INT} \left[ \frac{t(i) - t_{0}}{P}  \right],
     \label{eqphi}
   \end{equation}

\noindent where $t$ is the time, $t_{0}$ is the time origin, $P$ is a test period, and ${\rm INT}$ denotes the integer function. Because of the presence of observational gaps, it often happens that computed phases can have the same numerical values for several different test periods. Many periods are spurious because they correspond to gaps that may be present in the data \citep[e.g., daytime, seasons, etc.;][]{{1965ApJS...11..216L}}. If $P$ is the true period and $P_{\rm gaps}$ the period of gaps \citep{2007AJ....133.1470D}, spurious periods are given by

  \begin{equation}
     P_{\rm spur}^{-1} = P^{-1} \pm k \, P_{\rm gaps}^{-1}, 
     \label{palias}
   \end{equation}

\noindent where $k \in \mathbb{N}$. In some cases, a spurious period could be ranked as the best period.

We searched for periodic variable stars in our {\em initial catalog} (see Sect. 3.4) by applying various frequency analysis methods in combination. To search for the best period (or, equivalently, frequency), many methods require that a search range (given by limiting frequencies $f_{0}$ and $f_{N}$ at the low- and high-frequency ends, respectively) and resolution $\Delta f$ be specified first. Since we are dealing with data sets with various different time spans, we adapted the $f_0$ low-frequency limits to each light curve by  $f_0 = 0.5/T_{\rm tot}$, where $T_{\rm tot}$ is the total baseline of the observations, and we scaled the $\Delta f$ frequency resolution bas $\Delta f = 0.1/T_{\rm tot}$. The method described by \cite{1999A&AS..135....1E} was used to determine the high-frequency limit $f_N$.

\begin{figure*}[htbp]
   \centering

   \includegraphics[width=0.3\textwidth]{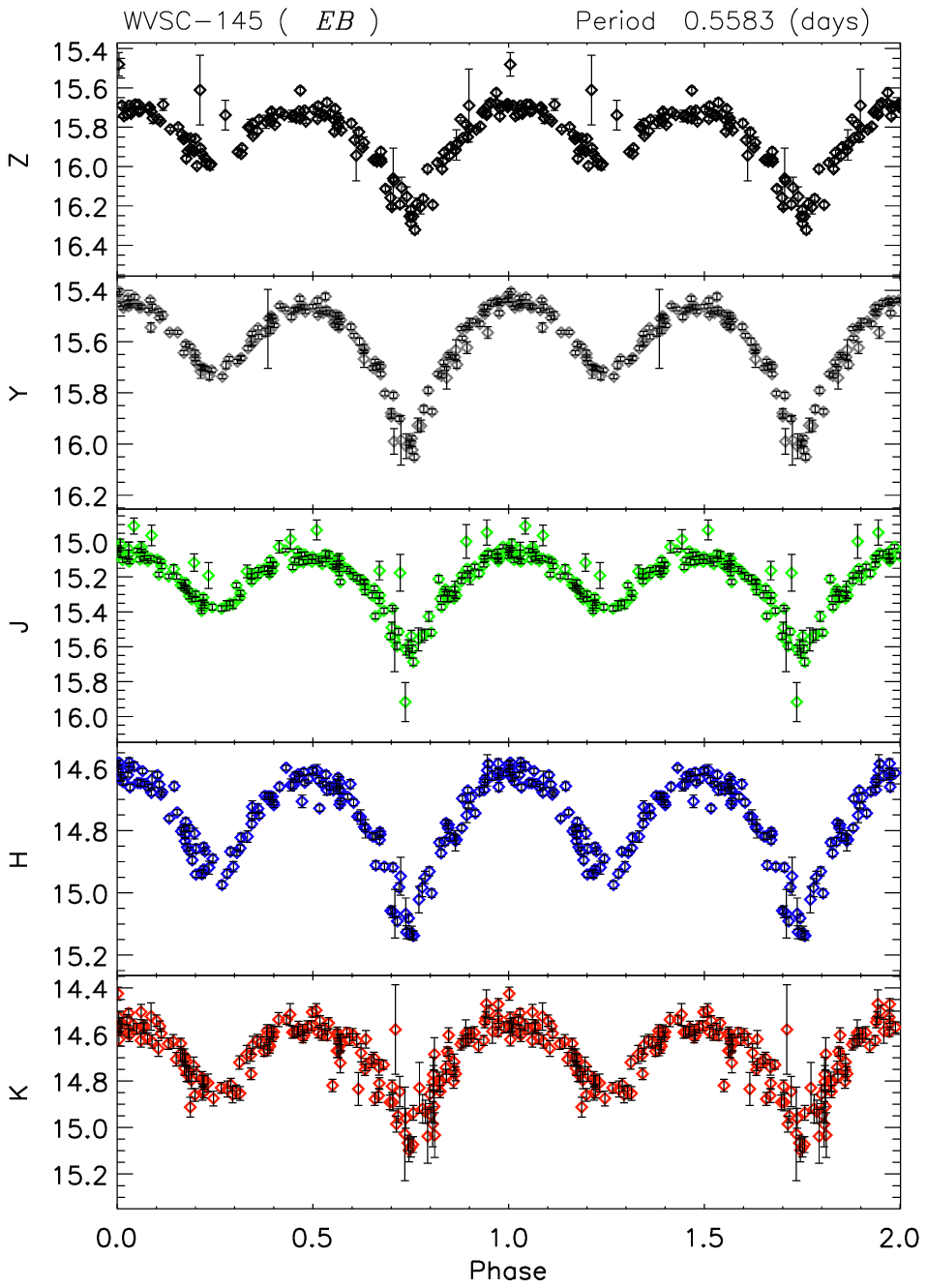} 
   \includegraphics[width=0.3\textwidth]{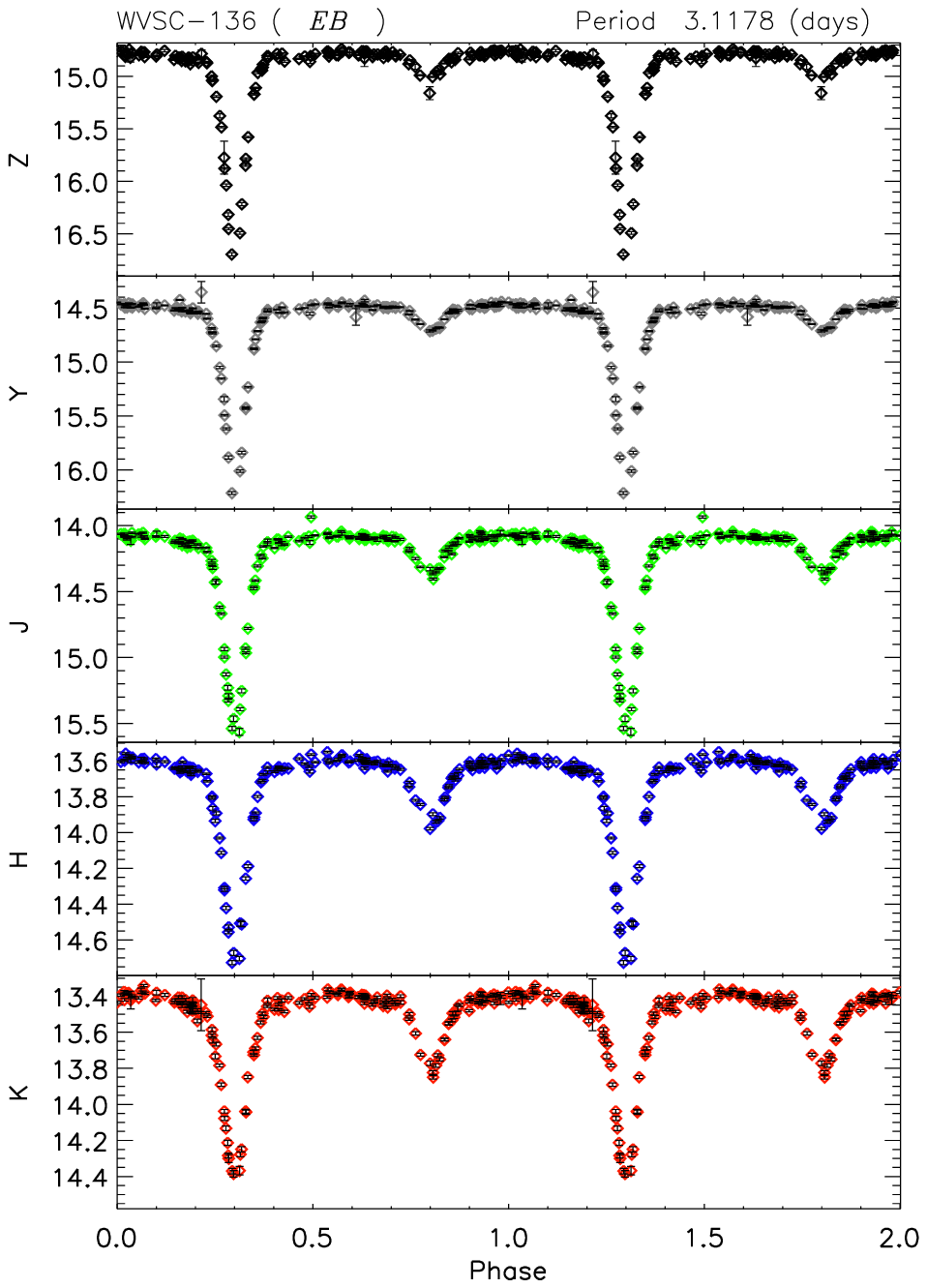} 
   \includegraphics[width=0.3\textwidth]{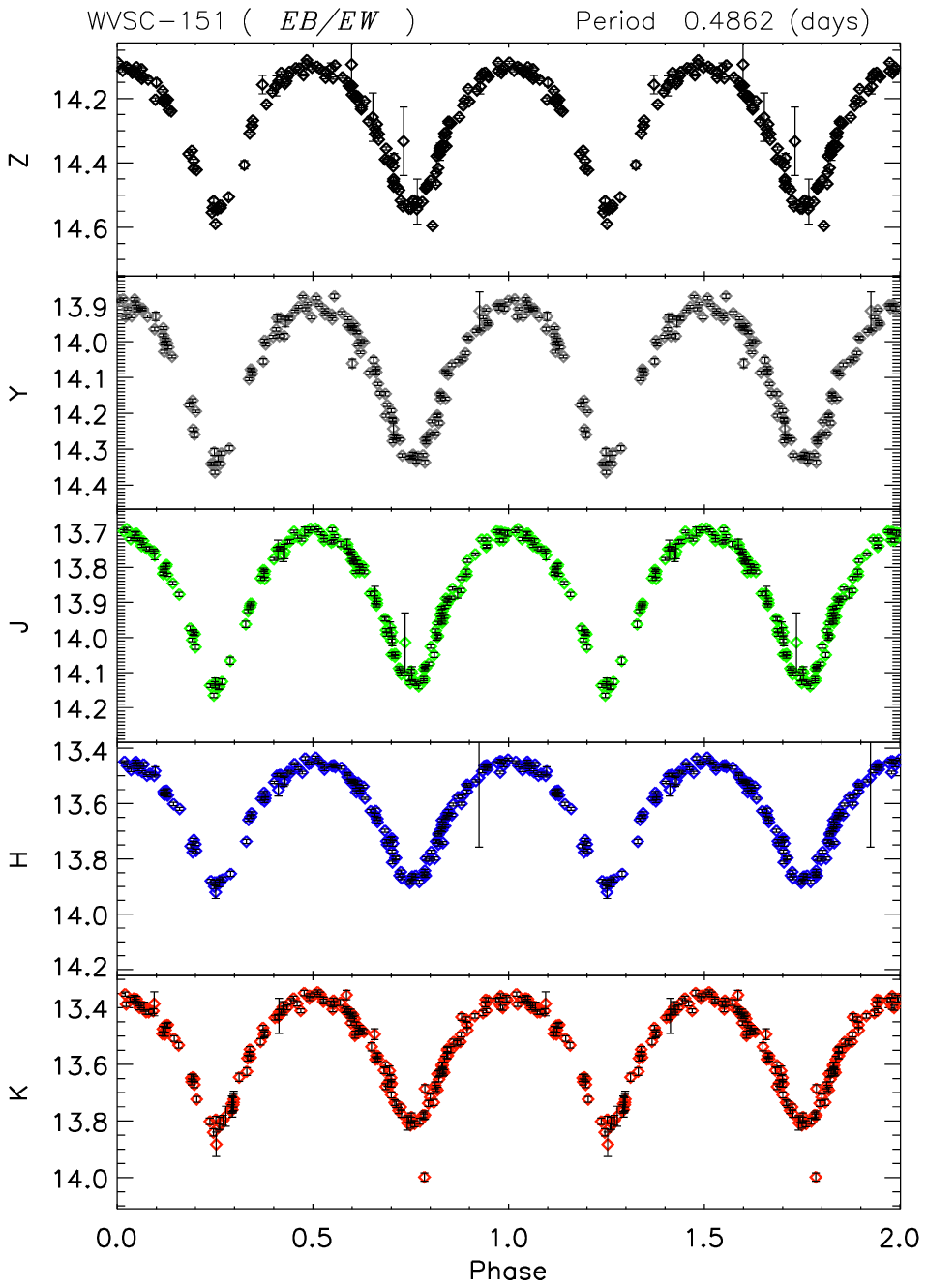}

   \includegraphics[width=0.3\textwidth]{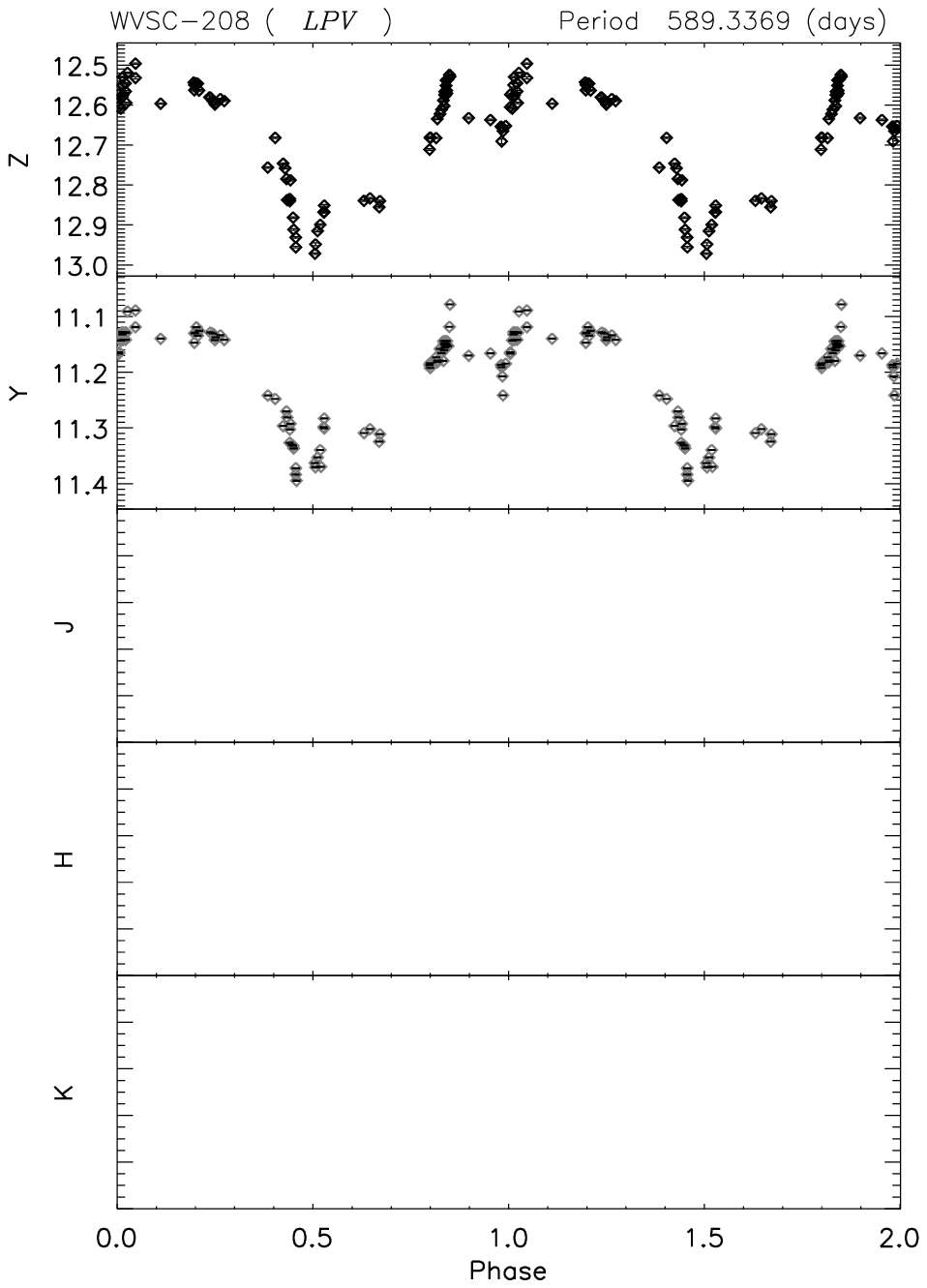} 
   \includegraphics[width=0.3\textwidth]{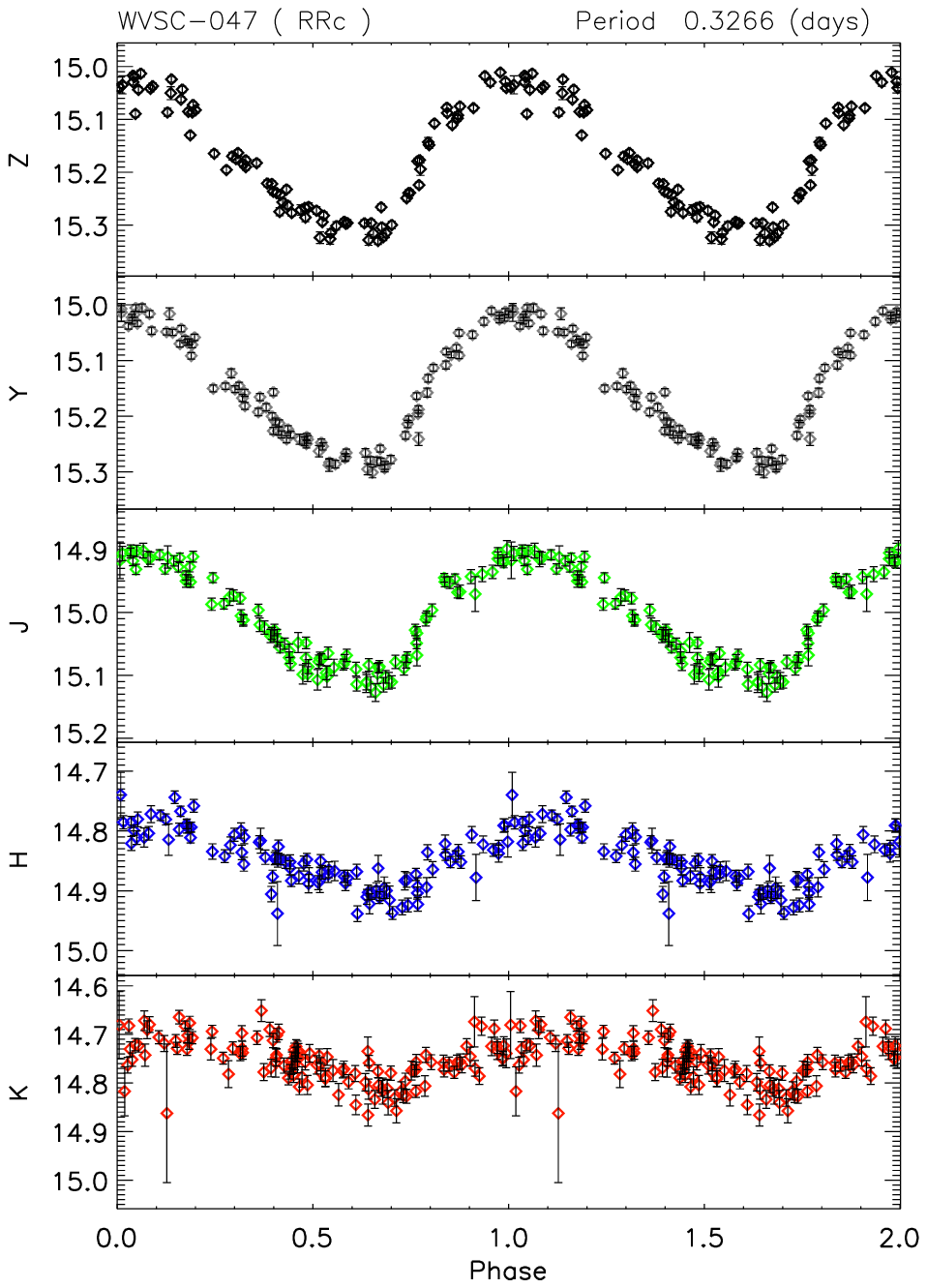} 
   \includegraphics[width=0.3\textwidth]{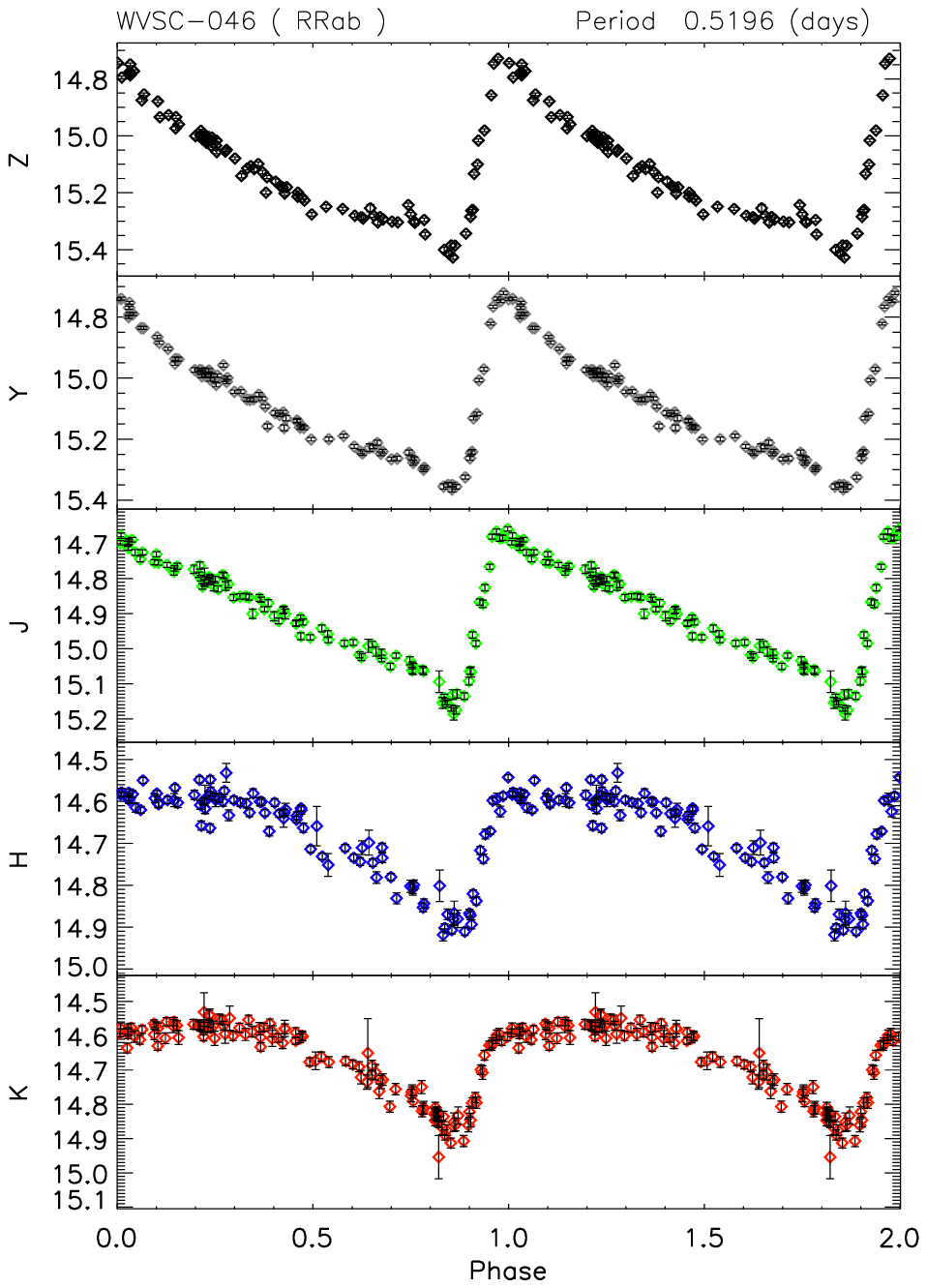} 

   \includegraphics[width=0.3\textwidth]{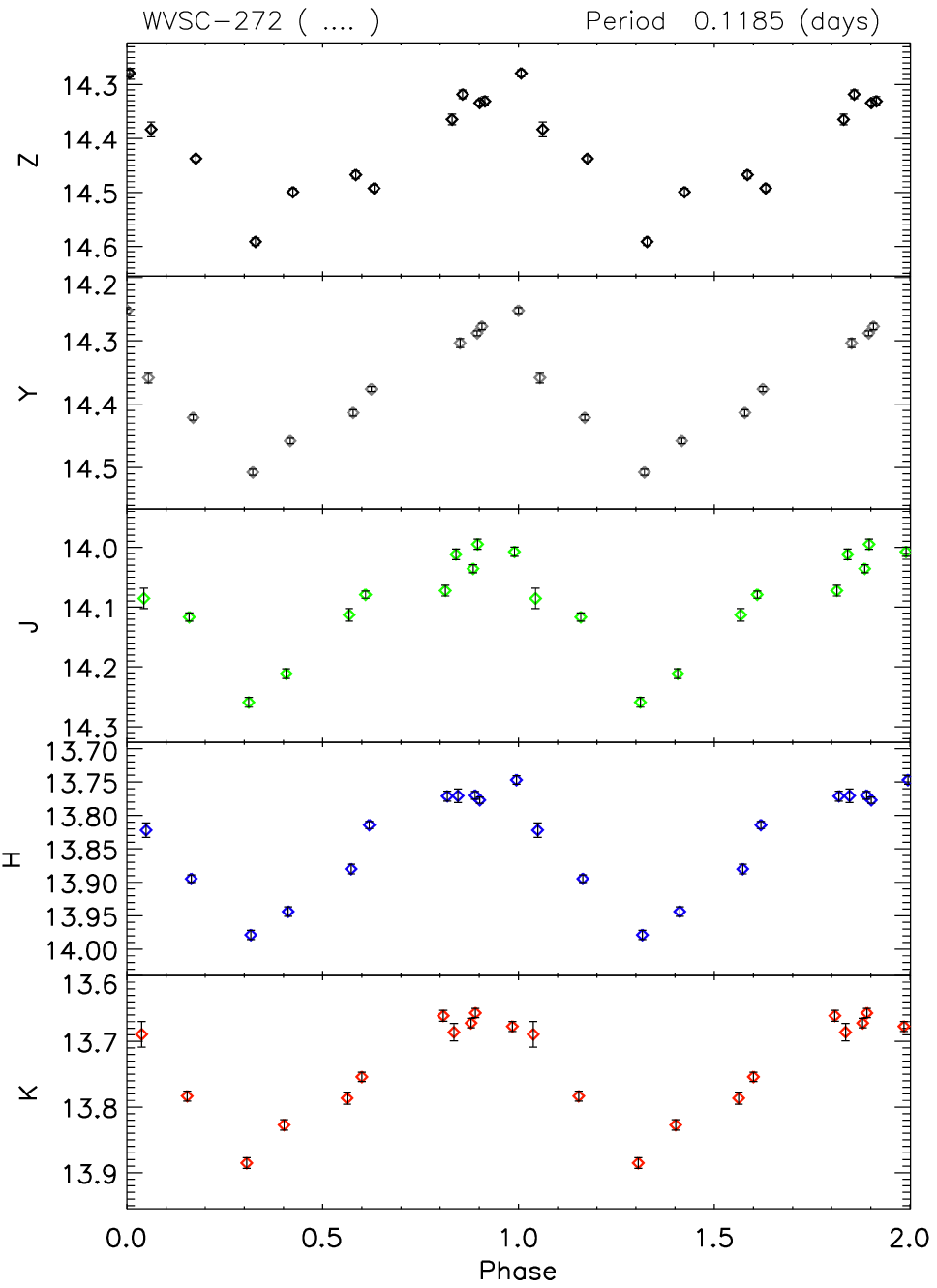} 
   \includegraphics[width=0.3\textwidth]{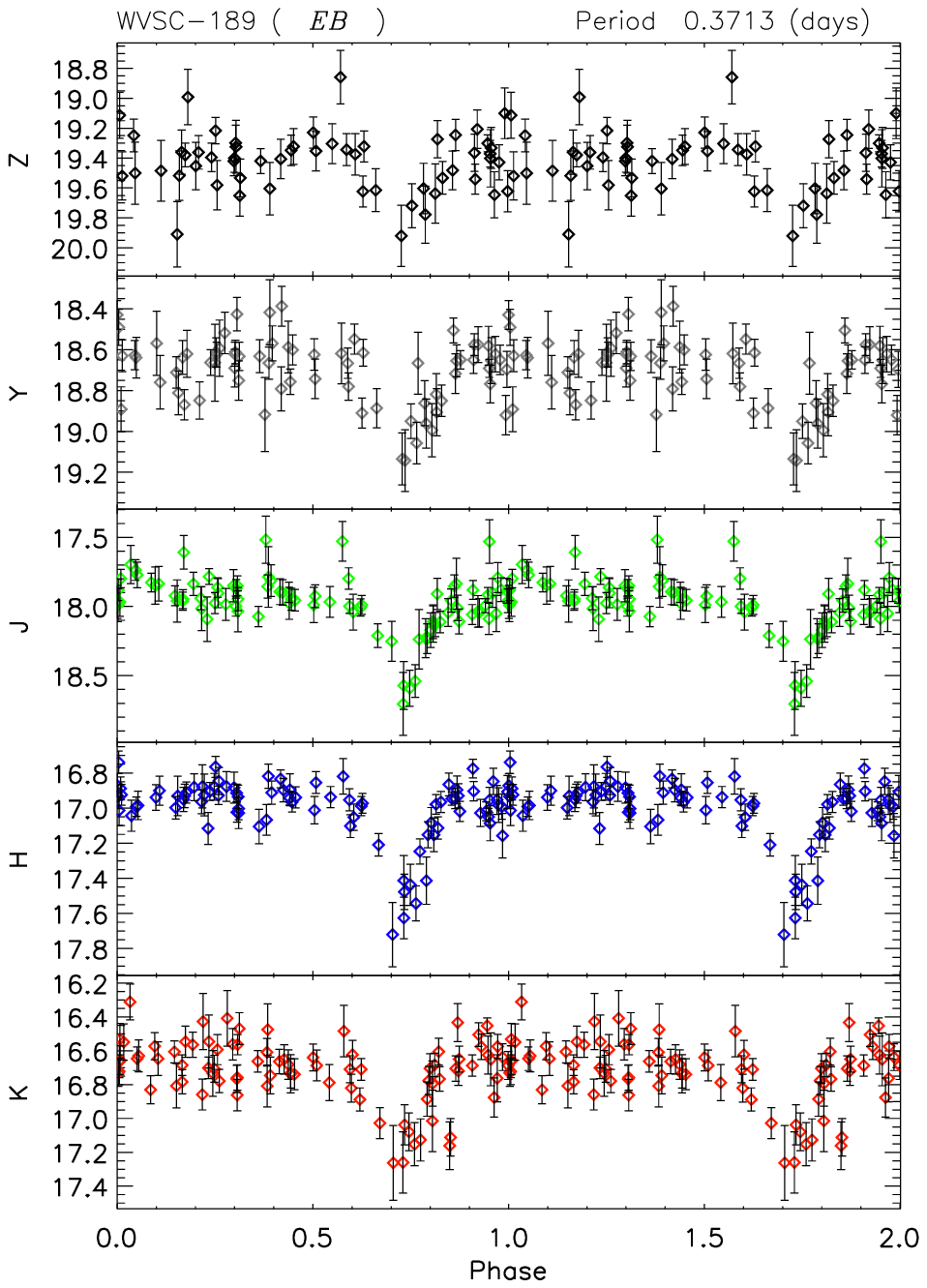} 
   \includegraphics[width=0.3\textwidth]{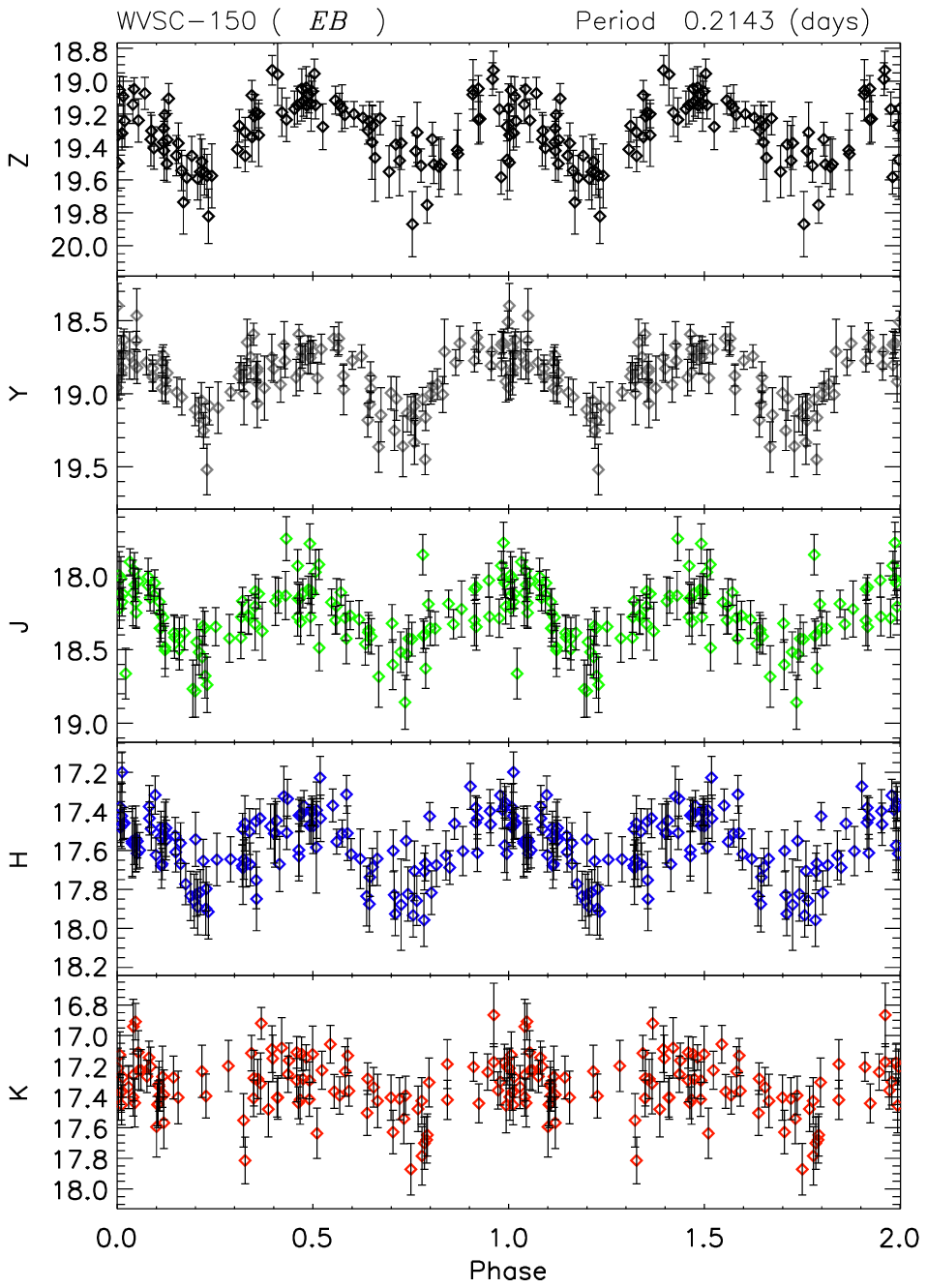}

   \caption{ Phase-folded light curves of selected objects from our catalog of periodic variables (C1), showing data in all five broadband filters of WFCAM. ID's, types,
  and periods for each object are shown in the headers. The object WVSC-208 was only detected in the $Z$ and $Y$ bands.}
   \label{lcbin02}
\end{figure*}

Initially, we applied the string-length minimization (SLM; \citealt{1965ApJS...11..216L}; \citealt{1996PASP..108..851S}) method on the light curves in our {\em initial catalog} of candidate variable stars. In this method, the period is found by minimizing the sum of the lengths of  the segments joining adjacent points in the phase diagrams (called the ``string lengths'')  using a series of trial periods, i.e.:

  \begin{equation}
     \Phi = \frac{\sum_{i = 1}^{N - 1} w_{i,i+1} \left| m_{i+1} + m_{i} \right|}{\sum_{i = 1}^{N - 1} w_{i,i+1} }, 
     \label{eqstet01}
   \end{equation}

\noindent where

  \begin{equation}
     w_{i,i+1} = \frac{1}{\left( \sigma_{i+1}^{2} + \sigma_{i}^{2} \right)\left( \varphi_{i+1} - \varphi_{i} + \epsilon \right)}, 
     \label{eqstet02}
   \end{equation}

\noindent and $\varphi_{i}$, $m_{i}$, and $\sigma_{i}$ are phases, magnitudes, and corresponding magnitude uncertainties, respectively, sorted in order of increasing phase according to the trial period. The variable $\epsilon$ is a term that is added to reduce the weight of very closely spaced data in phase space, which otherwise could have a weight approaching infinity \citep{1996PASP..108..851S}. In our work, we assumed $\epsilon = 1/N$, again following \citet{1996PASP..108..851S}.

To increase the probability that among the higher peaks of the periodogram derived on the basis of the SLM method the best period is included, we took the following additional steps: 

  \begin{itemize}

      \item  The SLM periodogram was independently computed for each broadband filter ($Y$, $Z$, $J$, $H$, and $K$), based on which all periods that presented a power greater than the $3\sigma$ level every band were selected; 
      \item  The SLM periodogram was also computed for the chromatic light curve, comprised of the sum of all broadband filters, and again all periods above the $3\sigma$ level were selected; 
      \item  The results of the previous two steps were then combined, yielding the best periods (i.e., the ones with the highest amplitude peaks) from both types of analysis.

  \end{itemize}

\noindent These steps are very important, because the same source can have photometry of high quality in some filters, but not others. 

In the second step of the period search, we applied three additional frequency analysis methods, to select the ten best among the periods selected by the SLM method. For this step, we used the ``classical'' Lomb-Scargle (LS) periodogram \citep{nl76,js82}, the generalized Lomb-Scargle (GLS) periodogram \citep{2009AA...496..577Z}, and the phase dispersion minimization \citep[PDM;][]{rs78} methods. We included both LS and GLS because the latter gives unbiased estimates of the standard deviation of the measurements only when the photometric errors are good, easily leading to spurious results otherwise. The PDM method was included due to its high sensitivity to highly non-sinusoidal variations with alternating peaks, as in the case of the light curves of eclipsing binary systems.


We proceeded with our analysis as follows:

\begin{itemize}
  \item For all light curves that were preselected by the SLM method, we independently computed the periodogram using each of these methods and for each filter separately;

  \item After inverting the spectra from the PDM and SLM methods, the periodograms for each filter were then normalized by the maximum power;
  
  \item A single, normalized power spectrum was then obtained for each method by averaging over all filters the results obtained in the previous step for each filter independently; 
  
  \item Finally, we obtained a ranked list of the best periods for each method.

\end{itemize}

The best periods should in principle be those that ranked highest in all four methods. However, not all methods rank periods in the same way. Therefore, to choose of periods in a more objective fashion, we computed a ``super rank'' index as the sum of the ranks provided by each of these five methods. We then selected the ten best periods as those ten most highly ranked, according to this new index.

Finally, in order to select the very best period, we use the $\chi^{2}$ test, in which Fourier coefficients $(a_{0}, a_{i}, b_{i})$ are derived that fit the data according to the following expression:

  \begin{equation}
     f(i) = a_{0} + \sum_{i = 1}^{n}\left\{ a_{i}\sin\left[2 k \pi \varphi(i) \right] + b_{i}\cos\left[2 k \pi \varphi(i) \right] \right\},  
     \label{eq_best_harm}
   \end{equation}

\noindent where $k = 1/P$. We used the Levenberg-Marquardt \citep{lev44,mar63} method to find the solutions and employed the statistical F-test to evaluate the results. We limited the amplitude of the fit to the observed magnitude range $(m_{\rm max}-m_{\rm min})$ and kept the number of harmonics to $n \leq 20$. The period that produced the lowest reduced $\chi^{2}$ value was selected as the true main period. Finally, we visually inspected of the phase diagrams, which narrowed down our selection to our {\em final catalog} containing 275 periodic variable stars, as described in more detail in the next section.

\section{Results and discussions}\label{sec:results}

\subsection{Catalog of periodic variable stars (C1)} \label{seccat01}

Proceeding as described in the previous section, we have thus obtained our final sample of 275 clearly periodic variable stars (C1). Their coordinates, periods, mean magnitudes, and the number of epochs in each filter are listed in Table 5. We also provide preliminary classifications for most of the newly discovered variables. The variability types of these sources were assigned by visual inspection of their phase diagrams, and  are also listed in Table 5, using the nomenclature of the General Catalog of Variable Stars\footnote{A full description of the nomenclature can be found at \url{http://www.sai.msu.su/gcvs/gcvs/iii/vartype.txt}} (GCVS, Samus 1977). Phase-folded light curves for some of these objects are shown in Figure \ref{lcbin02}.

\begin{figure*}[t]
    \includegraphics[width=9cm]{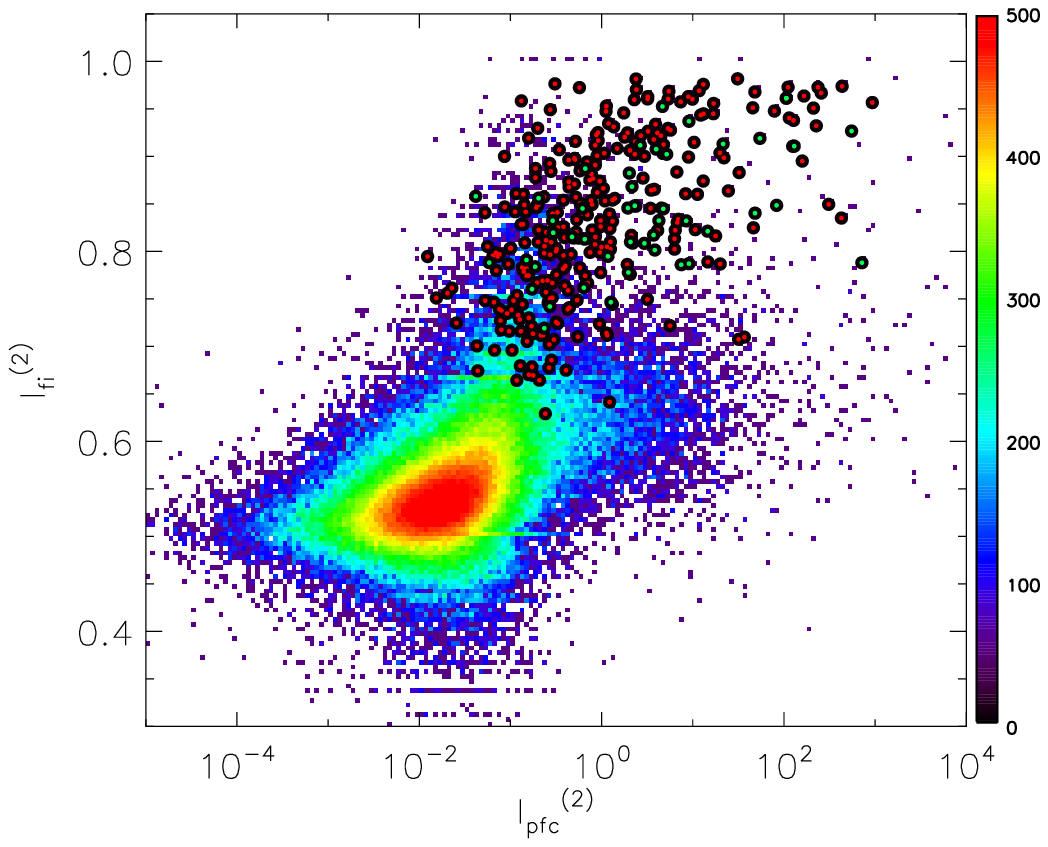}
    \includegraphics[width=9cm]{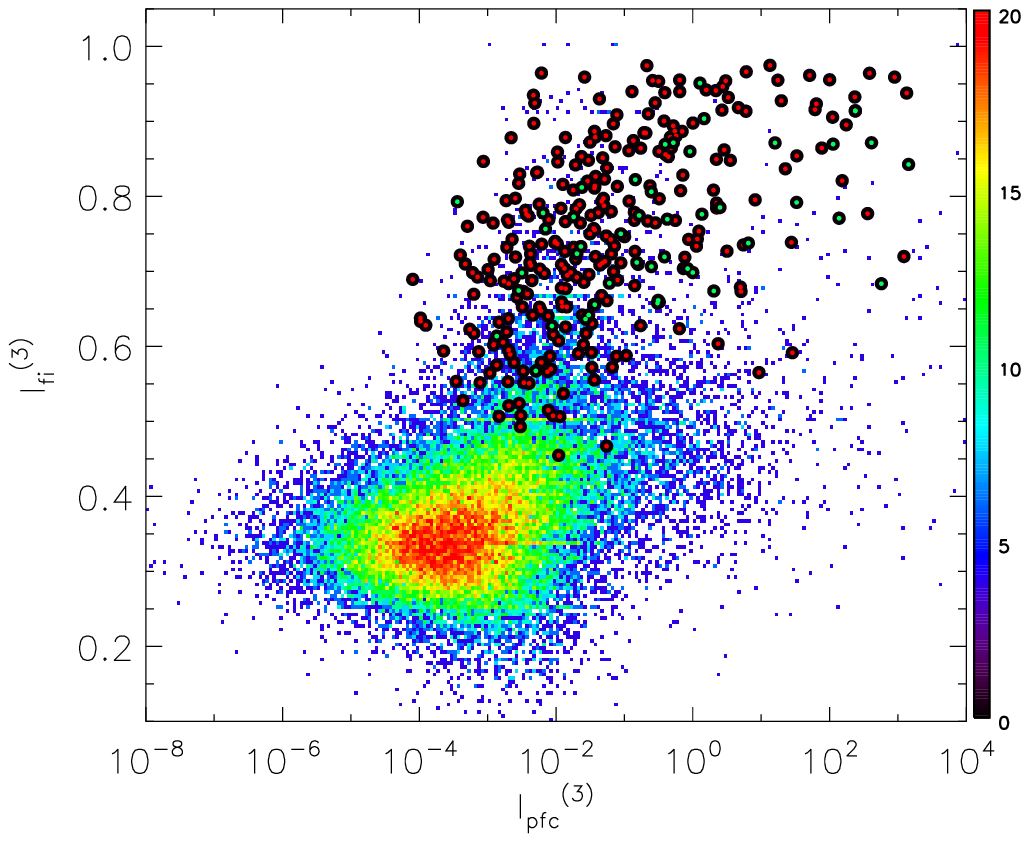}
  \caption{Distribution of $I_{\rm fi}$ versus $I_{\rm pfc}$ variability indices, for orders 2 ({\em left}) and 3 ({\em right}). The C1 and C2 sources are indicated by red and green circles, respectively.}
 \label{VIxVI}
\end{figure*}

\subsection{Catalog of aperiodic variable stars (C2)} \label{seccat02}

We also searched for those stars that, in spite of showing reasonably coherent light curves, do not show a clear main periodicity in the WFCAMCAL data, either because their variations are intrinsically aperiodic or because they have such long periods that these data were insufficient for deriving them. To identify such sources, we relied only on the variability indices, requiring that these have highly significant values, indicating the presence of correlated variations in the different WFCAM bandpasses. First, we selected sources by using a strong cutoff at the sampling- and magnitude-dependent 0.1\% significance level, equivalently to the procedure described in Sect.~\ref{sec_selecvar}. Table~\ref{tab_selecvar} shows the number of selected sources based on each of the cutoff surfaces shown in Figure~\ref{fig_indices}, before and after visual inspection. The number of sources selected by the $I_{\rm fi}^{(s)}$ index is less than a third of the sources selected using the $I_{\rm pfc}^{(s)}$ index, and at the same time, it includes 80\% of all sources in C1 (Sect.~\ref{seccat01}). This high efficiency at a relatively low false alarm rate favored the application of the $I_{\rm fi}^{(s)}$ index alone for selecting aperiodic variable candidates, which was followed by a visual inspection in order to reject likely false candidates.

Our procedure has led to an additional 44 sources, comprising our catalog of semi-regular or aperiodic variable stars and stars with uncertain periods (C2), shown in Table~\ref{tab-cat02}. The selected variable star candidates, including both periodic (C1) and non-periodic (C2) sources, are shown in the [$I_{\rm fi}^{(s)}$, $I_{\rm pfc}^{(s)}$] plane (for $s=2$ and 3) in Figure~\ref{VIxVI}. This figure clearly shows that the $I_{\rm fi}^{(s)}$ index is significantly more powerful, as far as distinguishing the variability in the WFCAM data is concerned, compared with the $I_{\rm pfc}^{(s)}$ index, since there is much less overlap between variable and non-variable sources in the former than the latter. As a word of caution, we note that the C2 catalog could still contain spurious sources, mainly due to sources that show correlated seasonal variations and/or correlated noise variations. Follow-up studies of these sources is thus strongly recommended, before conclusively establishing their variability status.

\begin{table}[htbp]
\caption{Number of variable star candidates selected by the different variability indices at different significance levels before and after (in parentheses) visual inspection.}
\centering                     
\begin{tabular}{l c c c c }     
\hline\hline
Sign. level & $I_{\rm pfc}^{(2)}$  & $I_{\rm pfc}^{(3)}$ & $I_{\rm fi}^{(2)}$ & $I_{\rm fi}^{(3)}$  \\ 
\hline               
 0.5\% & 4598(219) & 3574(192)  &  1292(242) & 1045(242) \\
 0.1\% & 1676(145) & 1337(141)  &  466(186) & 398(190) \\
\hline               
\end{tabular}
\label{tab_selecvar}
\end{table}

\begin{figure*}[htbp]
   \centering
   \includegraphics[width=9cm]{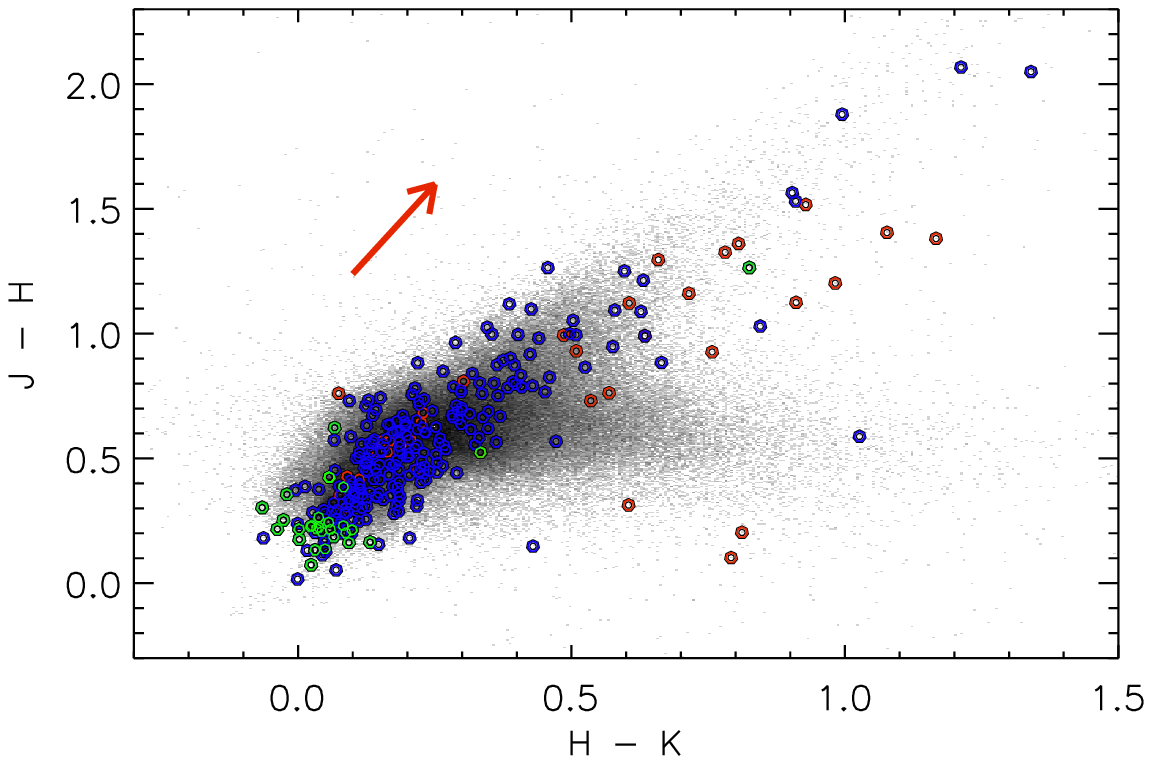} 
   \includegraphics[width=9cm]{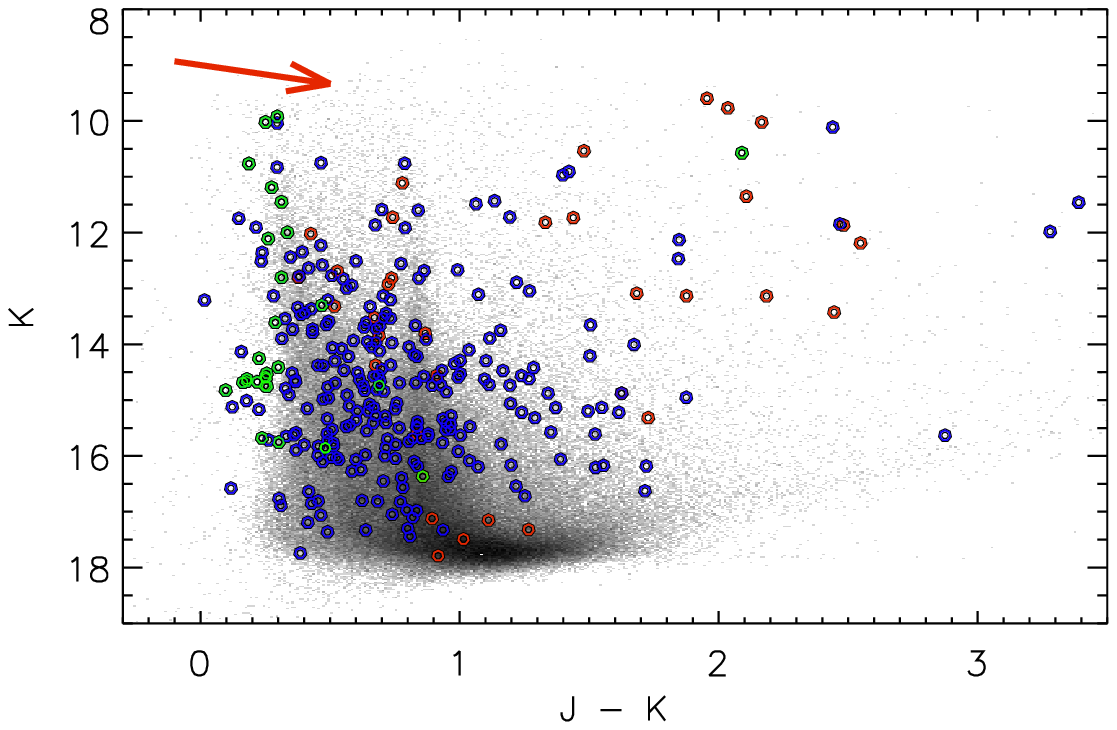} 
    \caption{Color-color {\em (on left panel)} and color-magnitude {\em (right panel)} diagrams of the analyzed sample. New objects in C1 (in blue) and C2 (in red) and previously known objects (in green) are shown with colored points. Red arrows indicate the reddening vectors.}
   \label{figcormag}
 \end{figure*}

\subsection{Cross-identifications} \label{cross}

As the final step in our analysis, we performed a systematic cross-check of the sample of 319 sources in C1 and C2 catalogs to identify previously known sources  and complement our catalog with data already in the literature. Among the cross-checked catalogs, one finds the SIMBAD database, the latest version of the General Catalog of Variable Stars \citep{nsea12}, the AAVSO International Variable Star Index (VSX v1.1, now including 284,893 variable stars; \citealt{2014yCat....102027W}), the New Catalog of Suspected Variable Stars \citep{ekea98}, and the Northern Sky Variability Survey (NSVS; \citealt{hoffman2009}) catalog, among many other databases incorporated in the International Virtual Observatory Alliance (IVOA), using the Astrogrid facility\footnote{\url{http://www.astrogrid.org/}}.

A delicate issue that we face when performing these extensive cross-checks between surveys with such diverse technical properties is their different astrometric accuracy. In this sense, we assumed a positional accuracy of $2 \arcsec$ in the sky coordinates for WFCAM, and then used optimized search radii according to the specific nature of each cross-matched database.  

Taking the distribution of our sources across the sky into account, along with the specific nature of the observations that comprise the WFCAMCAL catalog, which were aimed at observing standard star fields and hence tended to avoid very crowded regions, it did not come as a surprise that the cross-checking with variability surveys of the southern sky and Galactic central regions, such as OGLE, MACHO, and ASAS, did not result in any positive match. We extended the search further to other astronomical catalogs of non stellar and/or extragalactic objects (e.g., planetary nebulae, quasars, optical counterparts of GRBs, just to mention a few) and in different spectral bands (from radio to X-rays), but again finding no superpositions with our WFCAM sources.

At the end of this search, we found a total of 44 stars that were already known from previous studies. Among them, 37 sources are included in the VSX catalog, three of which are also GCVS objects (i.e., AM Tau, EH Lyn, UV Vir, which are an Algol-type eclipsing binary, a contact binary, and an ab-type RR Lyrae, respectively). The GCVS also lists five other eclipsing binaries and another RRab Lyrae (HM Vir).

The outer part of the globular cluster M3 (NGC~5272) is also partly covered by the WFCAMCAL pointings. Indeed, among our preselected variable candidates, we were able to recover 11 stars that had already been identified as RR Lyrae stars  by \citet{ccea05}, \citet{jjea12}, and \citet{2014yCat....102027W}. Cross-identifications and literature variability types of the previously known sources are given for C1 and C2 in Tables~\ref{tab-cat01} and \ref{tab-cat02}, respectively.

\subsection{Detection efficiency}

\begin{figure}
   \centering

   \includegraphics[width=0.22\textwidth]{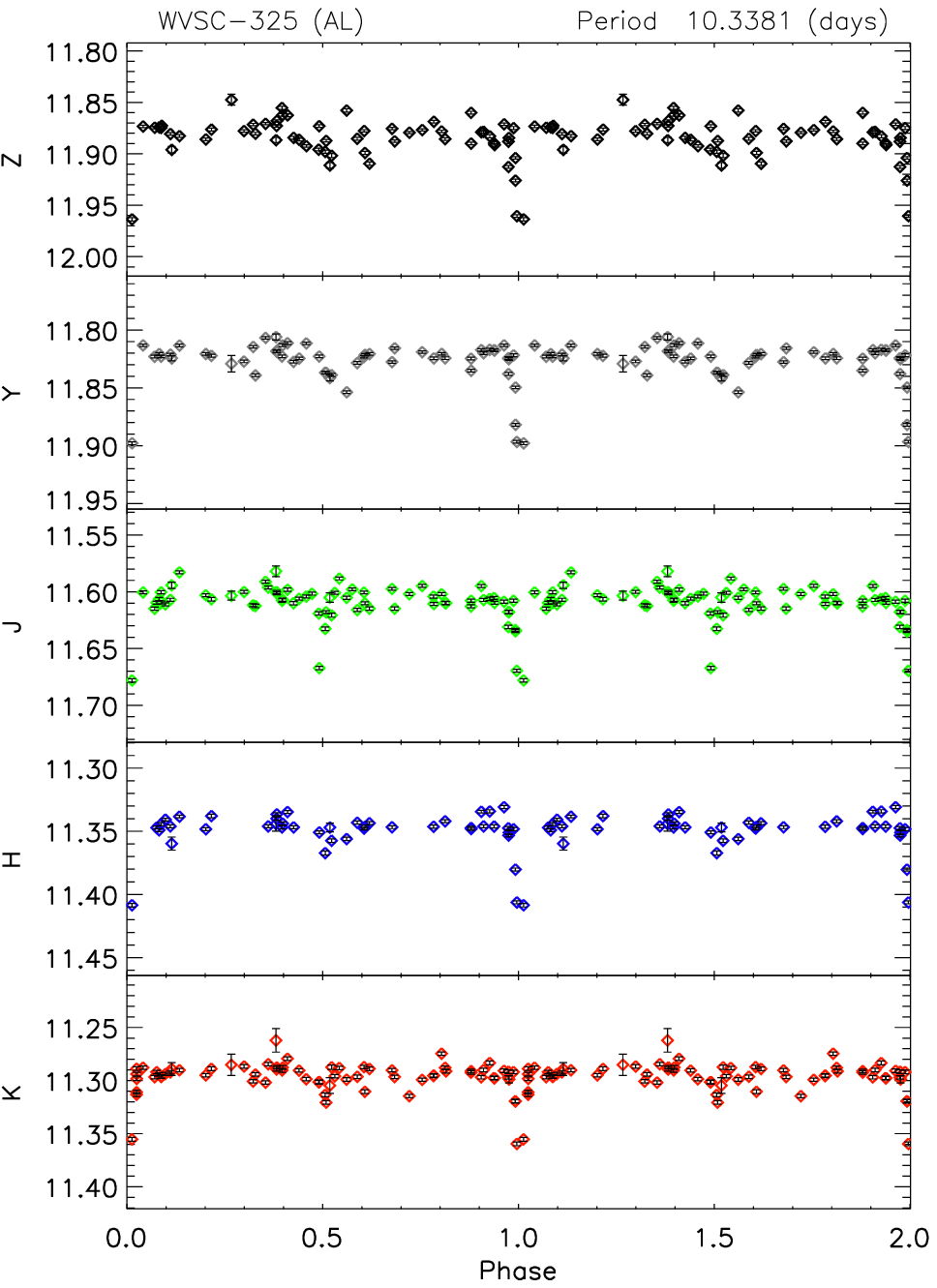} 
   \includegraphics[width=0.22\textwidth]{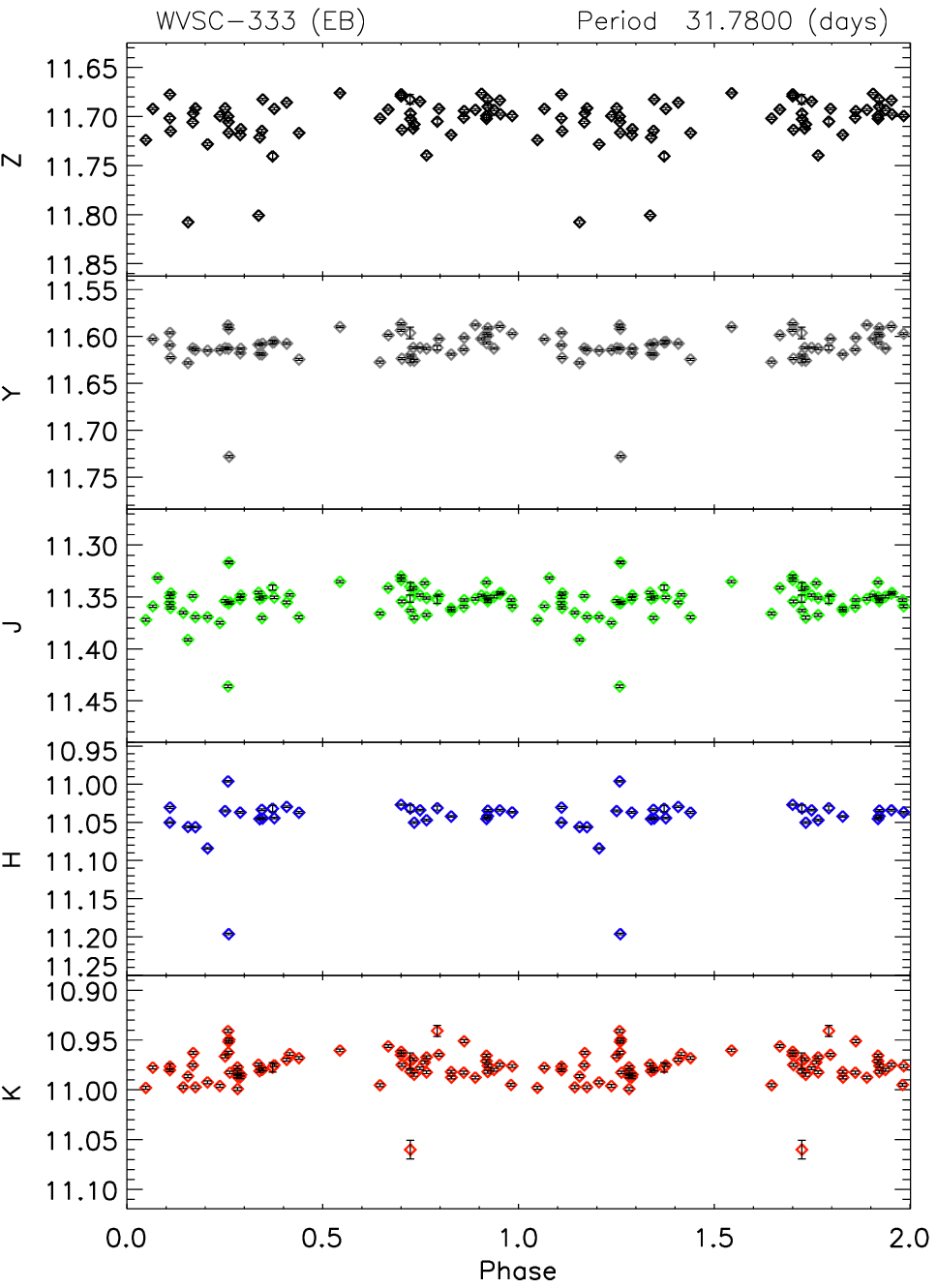} 

   \includegraphics[width=0.22\textwidth]{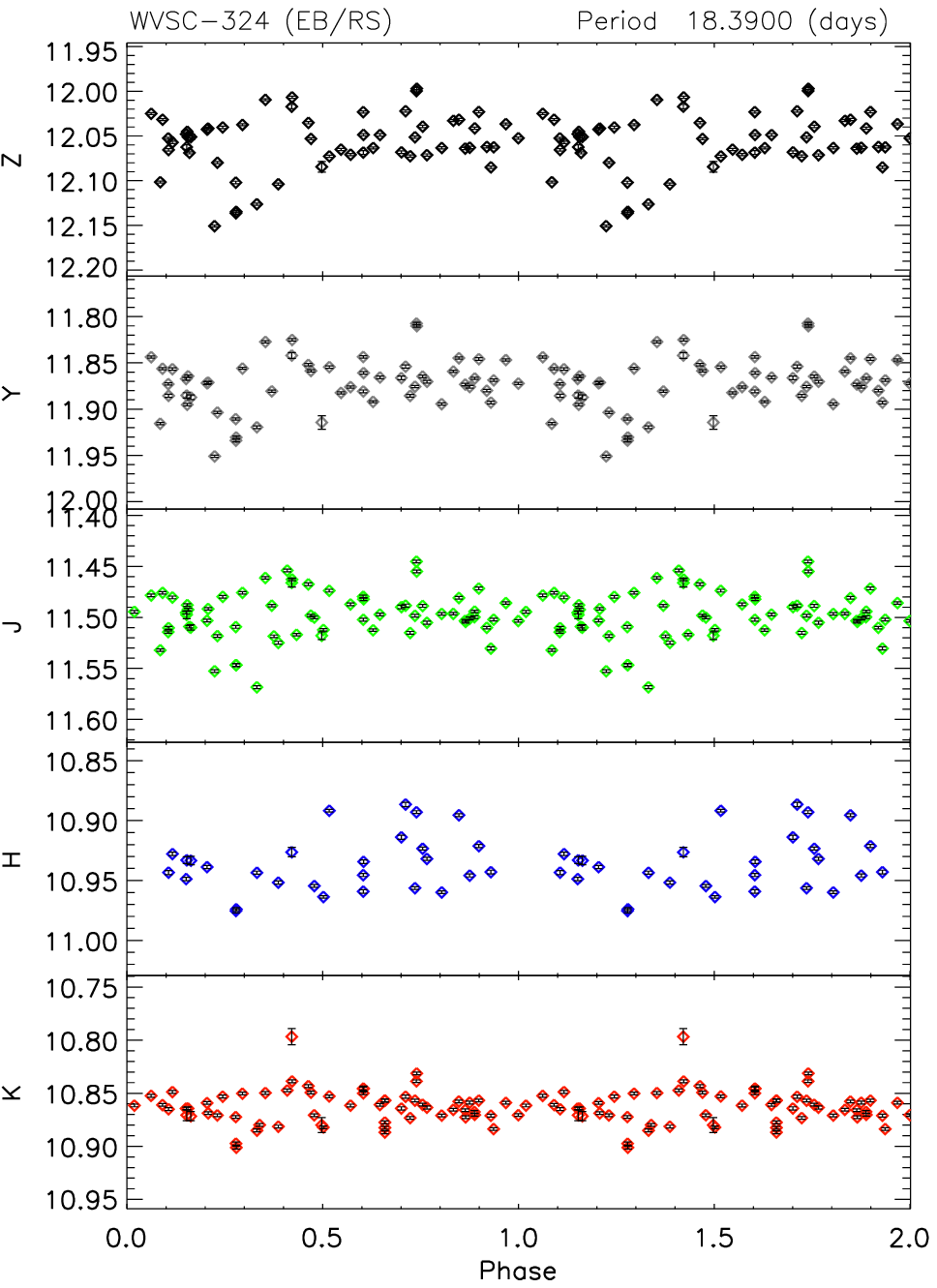}
   \includegraphics[width=0.22\textwidth]{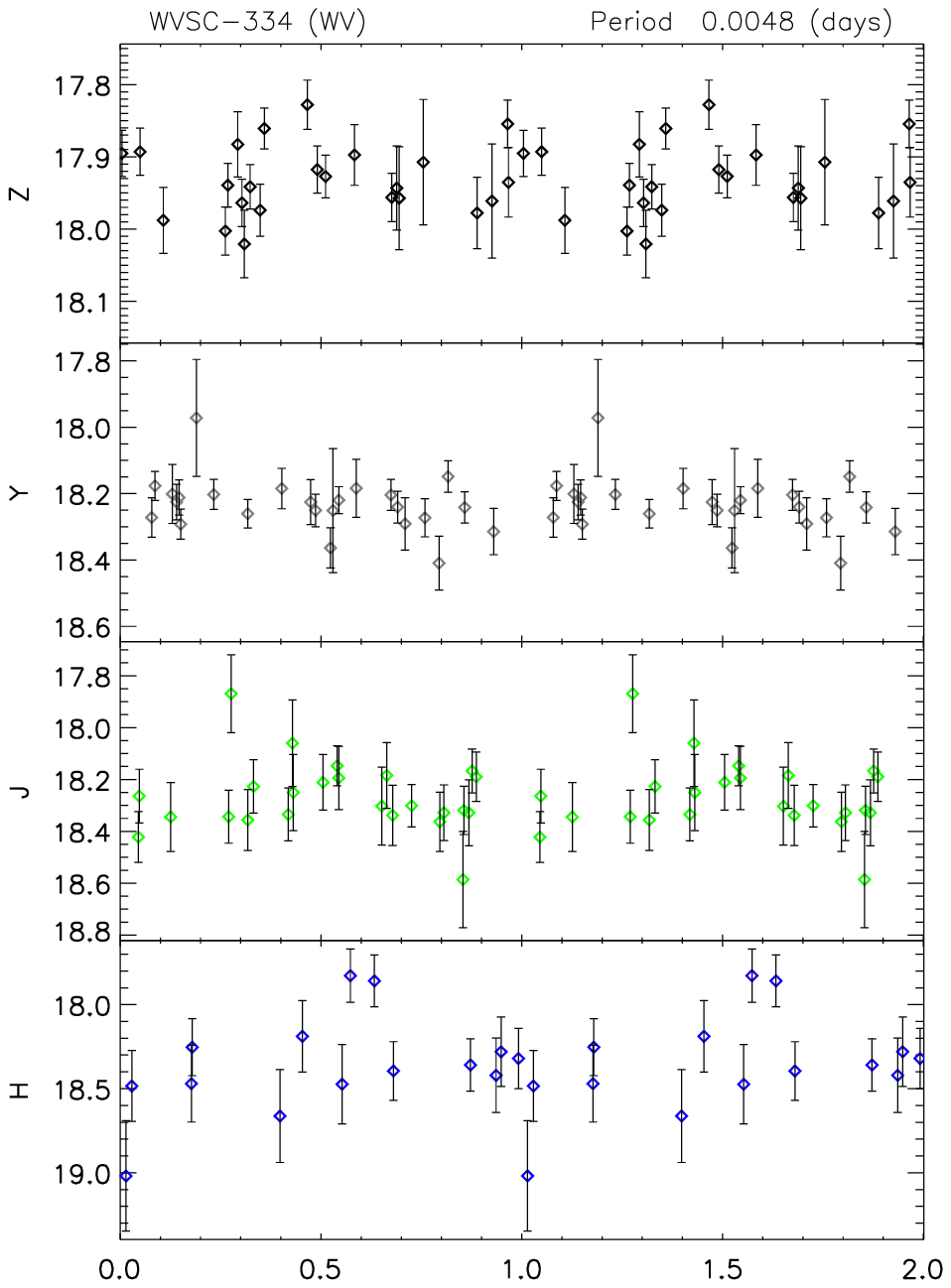}

   \caption{Example light curves of 4 previously known variable stars that were not detected in our analysis. ID's, variability types, and periods for each star are shown in the headers.}
   \label{lcs02}
\end{figure}

To estimate the variable star detection efficiency of our method, we determined whether any variable stars in the WFCAM calibration fields that are known from other catalogs were not detected by us. Similar to the procedure described in Sect.~\ref{cross}, we performed positional cross-matches of our \emph{initial database} of $216,\!722$ light curves with all existing survey catalogs of variable stars incorporated in the IVOA. We found a total of 15 known variables  (all of them periodic) that were missed by our search. Thirteen of them were excluded in the first broad selection phase (see Sect.~\ref{varindices}) owing to the low values of their variability indices. The properties of these stars are listed in Table \ref{tab-cat03}, and their WFCAM photometry is added to our catalog of periodic variable stars (C1), including a flag that refers to their non-detection as variables by our analysis.

Figure~\ref{lcs02} shows four typical examples among the non-detected variables. The non-detection of these sources is either due to the insufficient phase coverage of their magnitude variations by WFCAM data (primarily in the cases of long-periodic eclipsing binaries with very low fractional transit lengths, see Fig.~\ref{lcs02}, upper panels) or due to saturation (Fig.~\ref{lcs02}, lower left panel) or to a very low signal-to-noise ratio.
  
Since the heterogeneity and small number of objects known from other variable star catalogs overlap with the WFCAMCAL fields make them insufficient for a quantitative assessment of our detection efficiency, we performed further tests using synthetic data. To do this we first built a database of noise-free LCs as being harmonic fits (see Eq.~\ref{eq_best_harm}) of actual C1 data, in each filter. Next, we generated $10^5$ synthetic light curves, following the distributions of periods, amplitudes, and sampling of the final C1 catalog. Then,  these synthetic light curves were added to segments of the real light curves of non-variable stars from the WFCAMCAL database. Finally, we applied the same procedures of variability search and period analysis that we discussed in Sections~\ref{varindices} and \ref{secperiod} on the simulated data.

\begin{figure}
   \centering

   \includegraphics[width=0.45\textwidth]{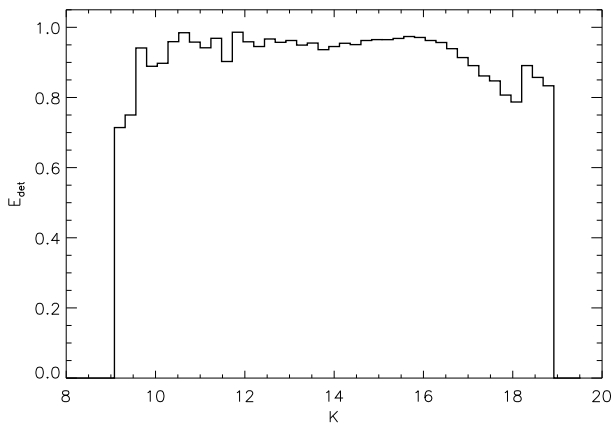} 

   \caption{ Detection efficiency $E_{det.}$ vs $K$ mean magnitude for $10^5$ synthetic variables.	}
   \label{effdiag}
\end{figure}

The result of our test is summarized by Figure~\ref{effdiag}, which shows the detection efficiency $E_{det.}=N_{det.}/N_{all}$ as a function of $K$ magnitude using bins of 0.25 mag, where $N_{det.}$ is the number of detected variables, and $N_{all}$ is the number of all variables, respectively. Based on this test, we estimate an average detection efficiency of $93\%$ over the complete magnitude range of the WFCAM database. Detection rates lower than $90\%$ are only present in the two extremes of the magnitude range, dominated by saturation in the bright end and low signal-to-noise ratio in the faint end. We note that the lower overall detection efficiency suggested by the catalog cross-matches is due to the biased magnitude distribution of the cross-matched sources toward the bright end of the WFCAM magnitude range.

\subsection{Photometric properties of the variables} \label{photprop}

Figure~\ref{figcormag} shows the variable and non-variable sources on the $(H-K) - (J-H)$ color-color and the $K - (J-K)$ color-magnitude planes. The variable sources cover the entire range of stellar parameter space in these planes. It is also clear from this figure that the area covered by the WFCAM Calibration fields have significant differential reddening, and a great fraction of the variables are reddened sources. This, together with the large aperture of the UKIRT telescope, explains the relatively low number of cross-identifications with previously known objects (Sect.~\ref{cross}): the WFCAMCAL database covers a range of faint NIR  magnitudes where most of the optical variability surveys cannot penetrate, while most of the known variables in those catalogs are too bright for WFCAM.

We found 32 highly reddened ($Z - K > 3$) variable sources, 17 of which are periodic. We found that the positions of these red sources are strongly clustered around the positions $\alpha=18^{\rm h} 29.5^{\rm m},~\delta=+1^{\circ}36'$ (25 sources), and $\alpha=7^{\rm h}0^{\rm m},~\delta=-4^{\circ}51'$ (8 sources). They are surrounded by a number of dark nebulae previously cataloged by \cite{2011PASJ...63S...1D} and \cite{2002A&A...383..631D}, respectively, suggesting that these sources might be embedded young stellar objects (see Tables~\ref{tab-cat01}, \ref{tab-cat02}).

\section{Conclusions}

In this paper, we have established the WFCAM Variable Star Catalog, based on a detailed analysis of the WFCAMCAL NIR database. Our catalog contains 319 variable point sources, among which are 275 that are clearly periodic, and it includes 44 previously known objects. All catalog entries including multiband light curve data are available online via the WFCAM Science Archive (WSA) \footnote{\url{http://surveys.roe.ac.uk/wsa/}}. Our approach to variability analysis included introducing a new, flux-independent variability index that is highly insensitive to the presence of outliers in the time-series data. A cross-matching procedure with previous variable star catalogs was also carried out, and the few sources with previous identification in the literature are noted. These catalogs represent one of the first such resource in the NIR, and thus an important first step toward the interpretation of future, more extensive NIR variability datasets, such as will be provided by the Vista Variables in the V\'{i}a L\'{a}ctea (VVV) Survey in particular \citep{mcea13}. A more detailed analysis of the different classes of variable stars detected in our catalog will be presented in a forthcoming paper.

\begin{acknowledgements}

Research activities of the Observational Astronomy Stellar Board of the Universidade Federal do Rio Grande do Norte are supported by continuous
grants from the CNPq and FAPERN Brazilian agencies. We also acknowledge financial support from INCT INEspaco/CNPq/MCT. CEFL acknowledges a CAPES
graduate fellowship and a CNPq/INEspaço postoctoral fellowship. Support for C.E.F.L., M.C., and I.D. is provided by the Chilean
Ministry for the Economy, Development, and Tourism's Programa Inicativa Cient\'ifica Milenio through grant IC\,12009, awarded to The
Millennium Institute of Astrophysics (MAS); by Proyectos FONDECYT Regulares \#1110326 and 1141141; by Proyecto Basal PFB-06/2007; and by Proyecto Anillo de Investigaci\'on  en Ciencia y Tecnolog\'ia PIA CONICYT-ACT 86. R.A. acknowledges support of 
Proyecto \mbox{GEMINI}-CONICYT \#32100022, of a PUC School of Engineering Postodoctoral Fellowship. 
Additional support by project VRI-PUC 25/2011 is also gratefully acknowledged. ICL acknowledges a CNPq/PNPD postdoctral fellowship.

\end{acknowledgements}


\onecolumn
\begin{longtab}
\begin{landscape}
\scriptsize
\begin{longtable}{l c c c c c c c c c c c c c c c c}

\caption{\label{tab-cat01} Periodic objects in the WFCAM Variable Star Catalog (C1)}\\
\hline\hline
ID [WSA] &  ID [WVSC] & ID & RA [deg.] & DEC [deg.] & type & P [d]  &  $\langle Z\rangle$ & $\langle Y\rangle$ & $\langle J\rangle$ & $\langle H\rangle$ & $\langle K\rangle$  & $N_Z$ & $N_Y$ &  $N_J$ &  $N_H$ &  $N_K$ \\
\hline
\endfirsthead
\caption{continued.}\\
\hline\hline
ID [WSA] &  ID [WVSC] & ID & RA [deg.] & DEC [deg.] & type & P [d]  &  $\langle Z\rangle$ & $\langle Y\rangle$ & $\langle J\rangle$ & $\langle H\rangle$ & $\langle K\rangle$  & $N_Z$ & $N_Y$ &  $N_J$ &  $N_H$ &  $N_K$ \\
\hline
\endhead
\hline
\multicolumn{17}{l}{Classifications taken from the literature and {\em our classifications} are marked by italics and normal fonts, respectively. }\\
\multicolumn{17}{l}{The abbreviations of variability types follow the same convention as the GCVS \citep{1997BaltA...6..296S}. Uncertain classifications are marked with `?'.}
\endfoot

858993461311  &  WVSC-001  &   UNC   &  146.283535  &  59.539435  &   UNC   &  0.158027  &  16.884  &  16.623  &  16.192  &  15.723  &  15.550  &  10  &  10  &  18  &  16  &    16 \\ 
858993464444  &  WVSC-002  &   UNC   &  145.560146  &  59.046595  &   \textit{RR}   &  0.49192  &  16.936  &  16.912  &  16.697  &  16.516  &  16.579  &  10  &  9  &  19  &  16  &    17 \\ 
858993475542  &  WVSC-003  &   CSS\_J033147.8+374651   &  52.949484  &  37.780864  &   EW   &  0.247579  &  15.796  &  15.522  &  15.013  &  14.385  &  14.188  &  59  &  79  &  91  &  84  &    93 \\ 
858993476434  &  WVSC-004  &   CSS\_J033109.1+375343   &  52.788116  &  37.895692  &   EW   &  0.286644  &  16.611  &  16.350  &  15.911  &  15.345  &  15.163  &  57  &  76  &  87  &  79  &     84 \\ 
858993477934  &  WVSC-005  &   UNC   &  53.345266  &  37.87284  &   UNC   &  0.127022  &  18.874  &  18.470  &  17.814  &  17.199  &  16.974  &  54  &  72  &  81  &  73  &    77 \\ 
858993478382  &  WVSC-006  &   UNC   &  53.424094  &  37.81507  &   UNC   &  2.4277  &  16.261  &  15.907  &  15.326  &  14.636  &  14.346  &  60  &  78  &  87  &  82  &   90 \\ 
858993479072  &  WVSC-007  &   UNC   &  53.53775  &  37.853985  &   \textit{EB?}   &  1.7298  &  18.447  &  18.008  &  17.327  &  16.652  &  16.370  &  55  &  74  &  83  &  75  &     81 \\ 
858993484033  &  WVSC-008  &   UNC   &  9.491549  &  37.781801  &   \textit{MP?}   &  10.205  &  16.733  &  16.561  &  16.078  &  15.633  &  15.412  &  99  &  102  &  115  &  114  &    117 \\
\end{longtable}
\end{landscape}
\end{longtab}



\begin{longtab}
\begin{landscape}
\scriptsize
\begin{longtable}{l c c c c c c c c c c c c c c c c}
\caption{\label{tab-cat02} Objects showing no main periodicity in the WFCAM Variable Star Catalog (C2).}\\
\hline\hline
ID [WSA] &  ID [WVSC] & ID & RA [deg.] & DEC [deg.] & type &   $\langle Z\rangle$ & $\langle Y\rangle$ & $\langle J\rangle$ & $\langle H\rangle$ & $\langle K\rangle$  & $N_Z$ & $N_Y$ &  $N_J$ &  $N_H$ &  $N_K$ \\
\hline
\endfirsthead
\caption{continued.}\\
\hline\hline
ID [WSA] &  ID [WVSC] & ID & RA [deg.] & DEC [deg.] & type &  $\langle Z\rangle$ & $\langle Y\rangle$ & $\langle J\rangle$ & $\langle H\rangle$ & $\langle K\rangle$  & $N_Z$ & $N_Y$ &  $N_J$ &  $N_H$ &  $N_K$ \\
\hline
\endhead
\hline
\multicolumn{16}{l}{Classifications taken from the literature and {\em our classifications} are marked by italics and normal fonts, respectively. }\\
\multicolumn{16}{l}{The abbreviations of variability types follow the same convention as the GCVS \citep{1997BaltA...6..296S}. Uncertain classifications are marked with `?'.}
\endfoot

858993544270  &  WVSC-276  &   UNC   &  9.60267  &  37.340321  &     UNC   &  19.896  &  19.621  &  19.041  &  18.353  &  -9.999  &  29  &  30  &  16  &  12  &     0 \\ 
858993598521  &  WVSC-277  &   UNC   &  250.220346  &  36.687414  &  UNC   &  17.654  &  17.434  &  17.224  &  16.700  &  16.361  &  69  &  73  &  82  &  75  &    83 \\ 
858993651900  &  WVSC-278  &   UNC   &  316.497648  &  30.537663  &  \textit{LPV?}   &  13.520  &  13.445  &  13.181  &  12.886  &  12.804  &  34  &  45  &  113  &  108  &   126 \\ 
858993762538  &  WVSC-279  &   UNC   &  205.382225  &  28.403026  &  UNC   &  14.304  &  14.082  &  13.652  &  13.081  &  12.927  &  72  &  80  &  102  &  100  &   126 \\ 
858993819409  &  WVSC-280  &   UNC   &  194.645106  &  21.684437  &  \textit{LPV?}   &  19.048  &  19.106  &  18.017  &  17.915  &  17.118  &  79  &  83  &  111  &  85  &   115 \\ 
858993855908  &  WVSC-281  &   UNC   &  88.255024  &  16.337922  &   UNC   &  14.392  &  14.174  &  13.836  &  13.409  &  13.319  &  37  &  42  &  42  &  42  &    55 \\ 
858993855917  &  WVSC-282  &   UNC   &  88.255236  &  16.41061  &    \textit{LPV?}   &  13.018  &  12.772  &  12.463  &  12.096  &  12.023  &  15  &  6  &  13  &  16  &    40 \\ 
858993939621  &  WVSC-283  &   UNC   &  312.800524  &  6.562438  &   \textit{LPV?}   &  14.810  &  14.617  &  14.194  &  13.669  &  13.523  &  140  &  142  &  161  &  158  &   169 \\
\end{longtable}
\end{landscape}
\end{longtab}

\begin{longtab}
\begin{landscape}
\scriptsize
\begin{longtable}{l c c c c c c c c c c c c c c c c}
\caption{\label{tab-cat03} Periodic objects in the WFCAM Variable Star Catalog non-detection as variables by our analysis (C3).}\\
\hline\hline
ID [WSA] &  ID [WVSC] & ID & RA [deg.] & DEC [deg.] & type & P [d]  &  $\langle Z\rangle$ & $\langle Y\rangle$ & $\langle J\rangle$ & $\langle H\rangle$ & $\langle K\rangle$  & $N_Z$ & $N_Y$ &  $N_J$ &  $N_H$ &  $N_K$ \\
\hline
\endfirsthead
\caption{continued.}\\
\hline\hline
ID [WSA] &  ID [WVSC] & ID & RA [deg.] & DEC [deg.] & type & P [d]  &  $\langle Z\rangle$ & $\langle Y\rangle$ & $\langle J\rangle$ & $\langle H\rangle$ & $\langle K\rangle$  & $N_Z$ & $N_Y$ &  $N_J$ &  $N_H$ &  $N_K$ \\
\hline
\endhead
\hline
\multicolumn{17}{l}{The missed sources in our broad selection are marked by italics.}\\
\multicolumn{17}{l}{The abbreviations of variability types follow the same convention as the GCVS \citep{1997BaltA...6..296S}.}
\endfoot

858993538871  &  \textit{WVSC-320}  &  2MASS J03314856+3723373  &  +52.9523380  &  +37.3937120  &  EA  &  11.778  &  10.549  &  10.454  &  10.297  &  10.107  &  10.018  &  8  &  7  &  19  &  10  &  21 \\ 
858993758877  &  \textit{WVSC-321}  &  Cl* NGC 5272 SAW V295  &  +205.4939460  &  +28.4663960  &  SX  &  0.036141  &  17.943  &  17.846  &  17.797  &  17.606  &  17.509  &  82  &  90  &  97  &  92  & 62 \\ 
858993852017  &  \textit{WVSC-322}  &  V0726 Tau  &  +88.0894190  &  +16.2985490  &  EB  &  1.984  &  10.778  &   -9.999  &  10.619  &  10.516  &  10.461  &  2  &  0  &  52  &  13  &  58 \\ 
858993875365  &  \textit{WVSC-323}  &  HT Cnc  &  +132.7801110  &  +11.8838720  &  EB  &  1.440  &  11.576  &   -9.999  &  11.171  &  10.769  &  10.620  &  3  &  1  &  4  &  3  &  10 \\ 
858993875597  &  WVSC-324  &  HU Cnc  &  +132.8056520  &  +11.8611550  &  EB/RS  &  18.390  &  12.057  &  11.872  &  11.497  &  10.936  &  10.862  &  56  &  55  &  65  &  28  &  61 \\ 
858993875889  &  \textit{WVSC-325}  &  HV Cnc  &  +132.8249810  &  +11.7650460  &  Al  &  10.338  &  11.885  &  11.827  &  11.609  &  11.349  &  11.296  &  49  &  48  &  67  &  42  &  67 \\ 
858993875909  &  \textit{WVSC-326}  &  Cl* NGC 2682 SAND 1045  &  +132.8292450  &  +11.8349220  &  EB  &  7.645  &  11.665  &  11.601  &  11.382  &  11.131  &  11.076  &  40  &  40  &  65  &  38  &  67 \\ 
858993875934  &  \textit{WVSC-327}  &  HX Cnc  &  +132.8320450  &  +11.8696420  &  RS  &  2.600  &  12.997  &  12.915  &  12.663  &  12.351  &  12.276  &  58  &  58  &  67  &  61  &  70 \\
\end{longtable}
\end{landscape}
\end{longtab}


\begin{thebibliography}{}
\bibitem[Alcock et al.(1993)]{1993ASPC...43..291A} Alcock, C., Allsman, R.~A., Axelrod, T.~S., et al.\ 1993, Sky Surveys.~Protostars to Protogalaxies, 43, 291 
\bibitem[Arnaboldi et al.(2007)]{maea07} 
  Arnaboldi, M., Neeser, M.~J., Parker, L.~C., et al.\ 2007, The Messenger, 127, 28
\bibitem[Arnaboldi et al.(2012)]{maea12} 
  Arnaboldi, M., Rejkuba, M., Retzlaff, J., et al.\ 2012, The Messenger, 149, 7  
\bibitem[Bakos et al.(2004)]{2004PASP..116..266B} Bakos, G., Noyes, R.~W., Kov{\'a}cs, G., et al.\ 2004, \pasp, 116, 266
\bibitem[Bla{\v z}ko(1907)]{sb07} 
  Bla{\v z}ko, S.\ 1907, Astron. Nachr., 175, 325
\bibitem[Blomme et al.(2011)]{2011MNRAS.418...96B} Blomme, J., Sarro, L.~M., O'Donovan, F.~T., et al.\ 2011, \mnras, 418, 96 
\bibitem[Cacciari et al.(2005)]{ccea05} 
  Cacciari, C., Corwin, T.~M., \& Carney, B.~W.\ 2005, \aj, 129, 267
\bibitem[Casali et al.(2007)]{casali2007} Casali, M., Adamson, A., Alves de Oliveira, C., et al.\ 2007, \aap, 467, 777 
\bibitem[Catelan et al.(2013)]{mcea13} 
  Catelan, M., Minniti, D., Lucas, P.~W., et al.\ 2013, in 40 Years of Variable Stars: A Celebration of Contributions by Horace A. Smith, p. 139 (arXiv:1310.1996) 
\bibitem[Cincotta et al.(1995)]{1995ApJ...449..231C} Cincotta, P.~M., Mendez, M., \& Nunez, J.~A.\ 1995, \apj, 449, 231 
\bibitem[Clausen et al.(2008)]{jcea08} Clausen, J.~V., Torres, G., Bruntt, H., et al.\ 2008, \aap, 487, 1095 
\bibitem[Cross et al.(2009)]{2009MNRAS.399.1730C} Cross, N.~J.~G., Collins, R.~S., Hambly, N.~C., et al.\ 2009, \mnras, 399, 1730 
\bibitem[Dalton et al.(2006)]{gdea06} Dalton, G.~B., Caldwell, M., Ward, A.~K., et al.\ 2006, \procspie, 6269, 62690X-1 
\bibitem[Damerdji et al.(2007)]{2007AJ....133.1470D} Damerdji, Y., Klotz, A., \& Bo{\"e}r, M.\ 2007, \aj, 133, 1470 
\bibitem[Deeming(1975)]{td75} Deeming, T.~J.\ 1975, \apss, 36, 137
\bibitem[Distefano et al.(2012)]{2012MNRAS.421.2774D} Distefano, E., Lanzafame, A.~C., Lanza, A.~F., et al.\ 2012, \mnras, 421, 2774 
\bibitem[Dobashi (2011)]{2011PASJ...63S...1D} Dobashi, K.\ 2011, \pasj, 63, 1
\bibitem[Drake et al.(2009)]{adea09} Drake, A.~J., Djorgovski, S.~G., Mahabal, A., et al.\ 2009, \apj, 696, 870 
\bibitem[Dutra \& Bica (2002)]{2002A&A...383..631D} Dutra, C.~M., \& Bica, E.\ 2002, \aap, 383, 631 
\bibitem[Dworetsky(1983)]{1983MNRAS.203..917D} Dworetsky, M.~M.\ 1983, \mnras, 203, 917 
\bibitem[Emerson et al.(2004)]{emerson2004} Emerson, J.~P., Irwin, M.~J., Lewis, J., et al.\ 2004, \procspie, 5493, 401 
\bibitem[Eyer(2006)]{2006ASPC..349...15E} Eyer, L.\ 2006, Astrophysics of Variable Stars, 349, 15 
\bibitem[Eyer \& Bartholdi(1999)]{1999A&AS..135....1E} Eyer, L., \& Bartholdi, P.\ 1999, \aaps, 135, 1 
\bibitem[Fruth et al.(2012)]{2012AJ....143..140F} Fruth, T., Kabath, P., Cabrera, J., et al.\ 2012, \aj, 143, 140 
\bibitem[Hambly et al.(2008)]{2008MNRAS.384..637H} Hambly, N.~C., Collins, R.~S., Cross, N.~J.~G., et al.\ 2008, \mnras, 384, 637 
\bibitem[Handler(2012)]{gh12} Handler, G.\ 2012, preprint (arXiv:1205.6407) 
\bibitem[Hewett et al.(2006)]{hewett2006} Hewett, P.~C., Warren, S.~J., Leggett, S.~K., \& Hodgkin, S.~T.\ 2006, \mnras, 367, 454 
\bibitem[Hoffman et al.(2009)]{hoffman2009} Hoffman, D.~I., Harrison, T.~E., \& McNamara, B.~J.\ 2009, \aj, 138, 466 
\bibitem[Hodgkin et al.(2009)]{Hodgkin2009} Hodgkin, S.~T., Irwin, M.~J., Hewett, P.~C.\ 2009, \mnras, 394, 675 
\bibitem[Irwin et al.(2004)]{2004SPIE.5493..411I} Irwin, M.~J., Lewis, J., Hodgkin, S., et al.\ 2004, \procspie, 5493, 411
\bibitem[Jurcsik et al.(2012)]{jjea12} Jurcsik, J., Hajdu, G., Szeidl, B., et al.\ 2012, \mnras, 419, 2173
\bibitem[Kaiser et al.(2002)]{nkea02} Kaiser, N., Aussel, H., Burke, B.~E., et al.\ 2002, \procspie, 4836, 154   
\bibitem[Kazarovets et al.(1998)]{ekea98} Kazarovets, E.~V., Samus, N.~N., \& Durlevich, O.~V.\ 1998, IBVS, 4655, 1
\bibitem[Krabbendam \& Sweeney(2010)]{2010SPIE.7733E..11K} Krabbendam, V.~L., \& Sweeney, D.\ 2010, \procspie, 7733  
\bibitem[Lafler \& Kinman(1965)]{1965ApJS...11..216L} Lafler, J., \& Kinman, T.~D.\ 1965, \apjs, 11, 216 
\bibitem[Levenberg(1944)]{lev44} Levenberg, K.\ 1944, Quarterly of Applied Mathematics, 2, 164
\bibitem[Lomb(1976)]{nl76} Lomb, N.~R.\ 1976, \apss, 39, 447    
\bibitem[Law et al.(2009)]{2009AAS...21346901L} Law, N.~M., Kulkarni, S., Ofek, E., et al.\ 2009, Bulletin of the   American Astronomical Society, 41, \#469.01
\bibitem[Lawrence et al.(2007)]{lawrence2007} Lawrence, A., Warren, S.~J., Almaini, O., et al.\ 2007, \mnras, 379, 1599 
\bibitem[Marquardt(1963)]{mar63} Marquardt, D. W. 1963, SIAM Journal of Applied Mathematics, 11, 431
\bibitem[McCommas et al.(2009)]{2009AJ....137.4707M} McCommas, L.~P., Yoachim, P., Williams, B.~F., et al.\ 2009, \aj, 137, 4707  
\bibitem[Minniti et al.(2010)]{dmea10} Minniti, D., Lucas, P.~W., Emerson, J.~P., et al.\ 2010, \na, 15, 433 
\bibitem[Perryman(2005)]{2005ASPC..338....3P} Perryman, M.~A.~C.\ 2005, Astrometry in the Age of the Next Generation of Large Telescopes, 338, 3 
\bibitem[Pojmanski(2002)]{2002AcA....52..397P} Pojmanski, G. \ 2002, \actaa, 52, 397
\bibitem[Pollacco et al.(2006)]{2006PASP..118.1407P} Pollacco, D.~L., Skillen, I., Collier Cameron, A., et al.\ 2006, \pasp, 118, 1407
\bibitem[Riello \& Irwin(2008)]{2008eic..work..581R} Riello, M., \& Irwin, M.\ 2008, 2007 ESO Instrument Calibration Workshop, 581  
\bibitem[Riess et al.(1998)]{area98} Riess, A.~G., Filippenko, A.~V., Challis, P., et al.\ 1998, \aj, 116, 1009   
\bibitem[Samus et al.(2012)]{nsea12} Samus N. N., Durlevich O. V., Kazarovets, E. V., et al. 2012, General Catalog of Variable Stars, VizieR On-line Data Catalog: B/gcvs 
\bibitem[Samus et al.(1997)]{1997BaltA...6..296S} Samus, N.~N., Durlevich, O.~V., \& Kazarovets, R.~V.\ 1997, Baltic Astronomy, 6, 296 
\bibitem[Scargle(1982)]{js82} Scargle, J.~D.\ 1982, \apj, 263, 835
\bibitem[Shin et al.(2012)]{2012AJ....143...65S} Shin, M.-S., Yi, H., Kim, D.-W., Chang, S.-W., \& Byun, Y.-I.\ 2012, \aj, 143, 65 
\bibitem[Siegel \& Majewski(2000)]{2000AJ....120..284S} Siegel, M.~H., \& Majewski, S.~R.\ 2000, \aj, 120, 284 
\bibitem[Simons \& Tokunaga(2002)]{simons+02} Simons, D.~A., \& Tokunaga, A.\ 2002, \pasp, 114, 169
\bibitem[Skrutskie et al.(2006)]{msea06} Skrutskie, M.~F., Cutri, R.~M., Stiening, R., et al.\ 2006, \aj, 131, 1163 
\bibitem[Stellingwerf(1978)]{rs78} Stellingwerf, R.~F.\ 1978, \apj, 224, 953   
\bibitem[Stetson(1981)]{pbs81} Stetson, P.~B.\ 1981, \aj, 86, 1500   
\bibitem[Stetson(1996)]{1996PASP..108..851S} Stetson, P.~B.\ 1996, \pasp, 108, 851 
\bibitem[Templeton(2004)]{mt04} Templeton, M.\ 2004, JAAVSO, 32, 41  
\bibitem[Tokunaga et al.(2002)]{tokunaga+02} Tokunaga, A.~T., Simons, D.~A., \& Vacca, W.~D.\ 2002, \pasp, 114, 180
\bibitem[Tonry et al.(2003)]{jtea03} Tonry, J.~L., Schmidt, B.~P., Barris, B., et al.\ 2003, \apj, 594, 1   
\bibitem[Udalski(2003)]{2003AcA....53..291U} Udalski, A.\ 2003, AcA, 53, 291 
\bibitem[Walker(2012)]{aw12} Walker, A.~R.\ 2012, \apss, 341, 43  
\bibitem[Watson et al.(2014)]{2014yCat....102027W} Watson, C., Henden, A.~A., \& Price, A.\ 2014, VizieR Online Data Catalog, 1, 2027 
\bibitem[Welch \& Stetson(1993)]{1993AJ....105.1813W} Welch, D.~L., \& Stetson, P.~B.\ 1993, \aj, 105, 1813 
\bibitem[Zechmeister \& K{\"u}lrster(2009)]{2009AA...496..577Z} Zechmeister, M. \& K{\"u}lrster, M.\ 2009, \aap, 496, 577 

\end{thebibliography}
\end{document}